\documentclass[sigconf,nonacm]{acmart}
\usepackage{xspace}

\AtBeginDocument{%
  \providecommand\BibTeX{{%
    Bib\TeX}}}

\usepackage{array}
\usepackage[caption=false,font=normalsize,labelfont=sf,textfont=sf]{subfig}
\usepackage{textcomp}
\usepackage{enumitem}
\usepackage{stfloats}
\usepackage{url}
\usepackage{verbatim}
\usepackage{wrapfig}
\usepackage{graphicx}
\def\BibTeX{{\rm B\kern-.05em{\sc i\kern-.025em b}\kern-.08em
    T\kern-.1667em\lower.7ex\hbox{E}\kern-.125emX}}
\usepackage{balance}

\usepackage{etoolbox}
\usepackage{dsfont}

\usepackage{amsmath,amssymb,amsfonts,bm}
\usepackage{graphicx}
\usepackage{textcomp}
\usepackage{xcolor}
\usepackage{layouts}
\usepackage[table]{xcolor}
\definecolor{SoftBlue}{RGB}{235, 242, 250} 
\definecolor{SageGreen}{RGB}{235, 245, 240} 

\usepackage{tabu}
\usepackage{textcomp}
\usepackage{xcolor}
\definecolor{commentgray}{RGB}{128,128,128} 
\usepackage{algorithm}
\usepackage{algpseudocode} 
 
\algrenewcommand\algorithmiccomment[1]{\textcolor{commentgray}{\textit{\(\triangleright\) #1}}}
\usepackage{subcaption}
\usepackage{url}
\usepackage{multirow}
\usepackage{framed}
\usepackage{adjustbox}
\usepackage{hyperref}
\usepackage{bbm}
\usepackage{xspace}
\usepackage{tcolorbox}
\tcbuselibrary{most}
\usepackage{marvosym}
\tcbuselibrary{most}

\usepackage{bbding} 
\usepackage{textcomp}
\usepackage{booktabs}
\usepackage{multirow}
\usepackage{bbm}
\usepackage{caption}
\usepackage{tcolorbox}
\usepackage{verbatim}
\usepackage{marginnote}
\usepackage{xspace}
\usepackage{soul}
\usepackage{siunitx}
\usepackage{xspace}
\usepackage{xspace}

\hypersetup{hidelinks,
	colorlinks=true,
	allcolors=black,
	pdfstartview=Fit,
	breaklinks=true}

\newtheorem{definition}{Definition}

\newcommand{\tool}{\textsc{SHARPEN}\xspace}
\newcommand{\commentout}[1]{}

\newcommand{\care}{CARE\xspace}
\newcommand{\ai}{AI-Lancet\xspace}
\newcommand{\ir}{IR\xspace}
\newcommand{\aprnn}{APRNN\xspace}
\newcommand{\inner}{INNER\xspace}
\newcommand{\prdnn}{PRDNN\xspace}
\newcommand{\adf}{ADF\xspace}
\newcommand{\idnn}{IDNN\xspace}
\newcommand{\deepr}{DeepRepair\xspace}
\newcommand{\fsgm}{FSGMix\xspace}
\newcommand{\rnnr}{ RNNRepair\xspace}
\newcommand{\fairn}{FairNeuron\xspace}
\newcommand{\neuronf}{NeuronFair\xspace}
\newcommand{\arachne}{Arachne\xspace}
\newcommand{\birdnn}{BIRDNN\xspace}
\newcommand{\vere}{VERE\xspace}
\newcommand{\nnr}{NNRepair\xspace}
\newcommand{\reassure}{REASSURE\xspace}
\newcommand{\autoric}{AutoRIC\xspace}
\newcommand{\airepair}{Airepair\xspace}
\newcommand{\piat}{PIAT\xspace}

\def\eqref#1{equation~(\ref{#1})}

\def\1{\bm{1}}

\DeclareMathAlphabet{\mathsfit}{\encodingdefault}{\sfdefault}{m}{sl}
\SetMathAlphabet{\mathsfit}{bold}{\encodingdefault}{\sfdefault}{bx}{n}

\newcommand{\dnns}{DNNs\xspace}
\newcommand{\dnn} {DNN\xspace}
\newcommand{\asr}{ASR\xspace}
\newcommand{\bsr}{BSR\xspace}

\begin{document}


\title{Shapley-Guided Neural Repair Approach via Derivative-Free Optimization}

\author{Xinyu Sun}
\affiliation{%
  \institution{National University of Defense Technology}
  \city{Changsha}
  \state{Hunan}
  \country{China}}
\email{sunxinyu17@nudt.edu.cn}

\author{Wanwei Liu}
\authornote{Corresponding author}
\affiliation{%
  \institution{ National University of Defense Technology}
  \city{Changsha}
  \state{Hunan}
  \country{China}}
\email{wwliu@nudt.edu.cn}

\author{Haoang Chi}
\affiliation{%
  \institution{ National University of Defense Technology}
  \city{Changsha}
  \state{Hunan}
  \country{China}}
\email{haoangchi618@gmail.com}

\author{Tingyu Chen}
\affiliation{%
  \institution{National University of Defense Technology}
  \city{Changsha}
  \state{Hunan}
  \country{China}}
\email{chentingyu@nudt.edu.cn}  

\author{Xiaoguang Mao}
\affiliation{%
  \institution{National University of Defense Technology}
  \city{Changsha}
  \state{Hunan}
  \country{China}}
\email{xgmao@nudt.edu.cn}

\author{Shangwen Wang}
\affiliation{%
  \institution{National University of Defense Technology}
  \city{Changsha}
  \state{Hunan}
  \country{China}}
\email{wangshangwen13@nudt.edu.cn}

\author{Lei Bu}
\affiliation{%
  \institution{Nanjing University}
  \city{Nanjing}
  \state{Jiangsu}
  \country{China}}
\email{bulei@nju.edu.cn}

\author{Jingyi Wang}
\affiliation{%
  \institution{ Zhejiang University}
  \city{Hangzhou}
  \state{Zhejiang}
  \country{China}}
\email{wangjyee@zju.edu.cn}

\author{Yang Tan}
\affiliation{%
  \institution{National University of Defense Technology}
  \city{Changsha}
  \state{Hunan}
  \country{China}}
\email{tyangty@163.com}

\author{Zhenyi Qi}
\affiliation{%
  \institution{National University of Defense Technology}
  \city{Changsha}
  \state{Hunan}
  \country{China}}
\email{qizhenyi@nudt.edu.cn}

\renewcommand{\shortauthors}{Sun et al.}

\begin{abstract}

Deep Neural Networks ({\dnn}s) are susceptible to a range of defects, including backdoors, adversarial attacks, and unfairness, which critically undermine their reliability.  Existing approaches mainly involve retraining, optimization, constraint-solving,  search algorithms, etc. However, most existing repair methods either rely on gradient calculations, restricting their applicability to networks with specific activation functions (e.g., ReLU), or use search algorithms whose localization and repair steps are often uninterpretable. Furthermore, these repair methods often lack generalizability across multiple properties.

  We propose \tool (SHApley-guided neural RePair via deriva-tive-freE optimizatioN), a framework that integrates an interpretable fault localization approach with a derivative-free optimization strategy. First, \tool introduces a  Deep SHAP-based white-box fault localization strategy that systematically quantifies the marginal contribution of each layer and neuron to erroneous outputs. In detail, we devise a hierarchical coarse-to-fine localization approach that reranks all layers by their aggregated impact and then locates faulty neurons or convolutional filters by analyzing activation divergences between property-violating and benign states. Subsequently,  \tool incorporates the Covariance Matrix Adaptation Evolution Strategy (CMA-ES) to repair the identified neurons. 
  CMA-ES leverages a covariance matrix to capture dependencies among variables, enabling gradient-free parameter search and coordinated adjustments across coupled neurons.
By combining interpretable fault localization with evolutionary optimization, \tool enables derivative-free repair across multiple architectures and is less sensitive to gradient anomalies and hyperparameters.

We demonstrate the effectiveness and efficiency of \tool on three repair tasks: backdoor removal, adversarial mitigation, and unfairness repair.  
Effectively balancing property repair and accuracy preservation, \tool outperforms state-of-the-art baselines by up to 10.56\% in backdoor removal (over INNER), 5.78\% in adversarial mitigation (over CARE), and 11.82\% in unfairness repair (over IDNN).
Notably, \tool is capable of handling diverse repair tasks, and its modular design is seamlessly plug-and-play with different derivative-free optimizers in the repair stage, highlighting its flexibility across the evaluated settings.

\end{abstract}



\maketitle 

\section{Introduction}

With the widespread adoption of Deep Neural Networks (DNNs) into safety-critical systems (e.g., autonomous vehicles, medical diagnostics), ensuring their reliability and trustworthiness has become a paramount priority \cite{DBLP:journals/corr/BojarskiTDFFGJM16,vieira2017using}. While these models demonstrate remarkable capabilities, they are susceptible to erroneous behaviors that can lead to severe consequences. These defects manifest in various forms, including backdoor attacks, where hidden triggers cause targeted misclassifications \cite{DBLP:conf/ccs/YaoLZZ19}; adversarial vulnerabilities, where imperceptible perturbations fool the model~\cite{DBLP:conf/sp/Carlini017}; and unfairness, where models exhibit discriminatory behavior against certain demographic groups \cite{DBLP:journals/pacmpl/Bastani0S19}. Existing verification and testing techniques can effectively detect such flaws, but they cannot repair them \cite{yang2021enhancing,DBLP:conf/qrs/ShenWC18}. Distinct from traditional software verification, neural network repair is a critical link in the development loop. While conventional software repair addresses logical inconsistencies, neural network repair operates at a numerical level, inherently lacking the interpretability found in symbolic logic. Therefore, developing interpretable methods to repair these defects in pre-trained models is a critical and pressing challenge for ensuring their safe deployment.

In response to these challenges, the research community has proposed various neural network repair strategies. The overarching goal is to eliminate undesired behaviors while preserving the model's ability on specific tasks and avoiding complete and costly retraining. Existing approaches can be categorized into three types based on their technical mechanisms. One major category involves search-based methods, which utilize heuristic optimization algorithms like Particle Swarm Optimization (PSO) \cite{DBLP:conf/icnn/KennedyE95} or evolutionary strategies to search for optimal weight adjustments for a set of identified faulty neurons, such as \care 
 \cite{DBLP:conf/icse/Sun0PS22} and Arachne \cite{DBLP:journals/tosem/SohnKY23}. Another category relies on constraint solving and formal methods \cite{DBLP:series/faia/BarrettSST09, DBLP:conf/stoc/Karmarkar84, DBLP:books/cu/BV2014}. These techniques strictly formalize the repair task to achieve provably correct patches. For instance, \prdnn \cite{DBLP:conf/pldi/SotoudehT21} introduces a novel architecture to reduce the repair problem to a solvable Linear Programming (LP) problem, a strategy also explored by other works like \aprnn \cite{DBLP:journals/pacmpl/TaoNMT23}. A third category focuses on direct structural modification, where the repair is achieved by altering the network's information flow rather than optimizing the model weights. For example, the \idnn 
 \cite{DBLP:conf/issta/ChenWS0024} adopts a technique that isolates and ``freezes'' the outputs of faulty neurons.

Despite these advances, a trade-off between scalability and precision in current methods limits their practical applicability. Approaches based on formal methods and constraint solving, while providing theoretical guarantees, often struggle to scale to the larger and more complex neural networks and may be limited to specific activation functions or properties \cite{DBLP:conf/iclr/FuL22,DBLP:journals/tosem/SunLWCTM25, DBLP:journals/corr/abs-2508-08151, DBLP:conf/icse/LiCZZ0LGL24, DBLP:conf/issta/DasuKTT24, DBLP:journals/pacmpl/TaoNMT23, DBLP:conf/pldi/SotoudehT21, DBLP:conf/aaai/FuWZWFH0C024, DBLP:conf/cav/UsmanGSNP21}. Conversely, many search-based and heuristic methods can handle complex models but may require access to internal states, such as gradients, or lack a precise and interpretable way to locate the root cause of a defect, leading to inefficient search or suboptimal repairs. These issues require a more reliable repair framework that combines precise fault localization with effective repair optimization across multiple repair settings.

\begin{figure}[t]
  \centering
   \includegraphics[width=1.0\columnwidth]{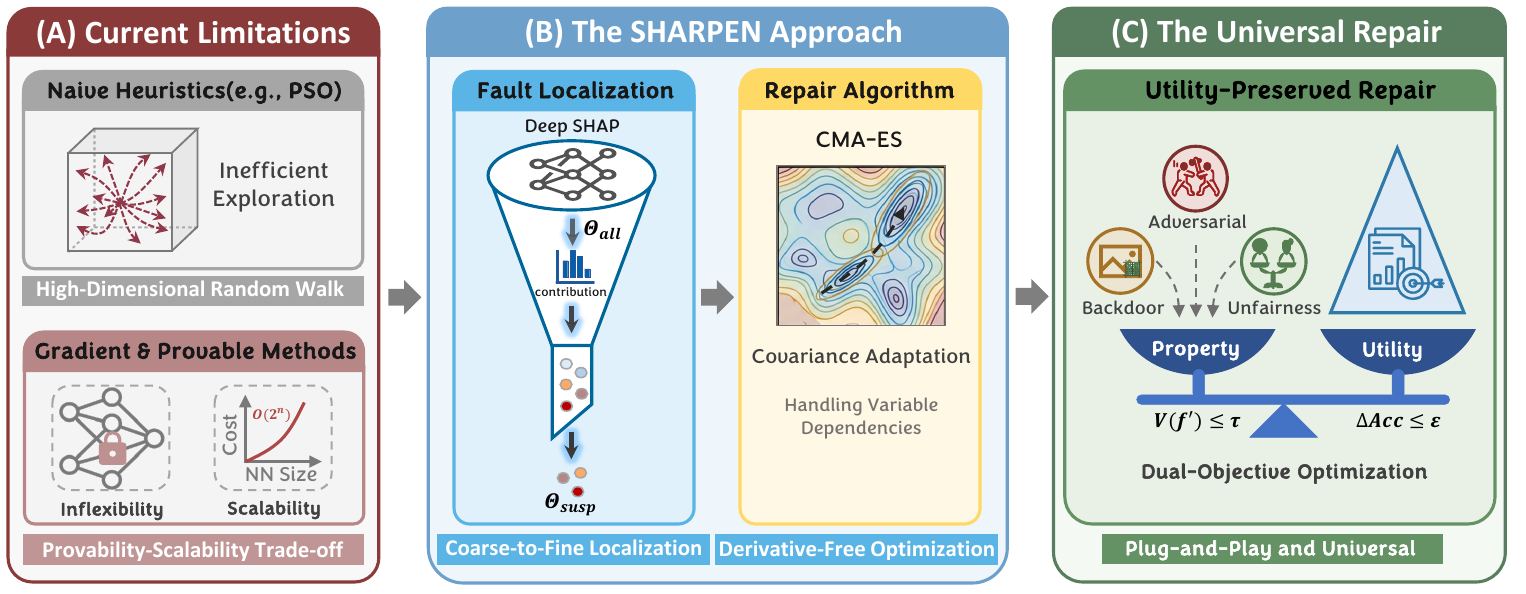}   
  \caption{\textbf{Motivation of \tool.} 
\textbf{(A) Dilemma:} Methods trade off between blind exploration and unscalable rigid constraints.
\textbf{(B) Insight:} Recognizing that neural network faults often suffer from sparse critical neurons, we leverage interpretability to constrain the search space, transforming high-dimensional optimization into a local problem.
\textbf{(C) Goal:} Achieving \textit{utility-preserved repair}, satisfying the property ($V(f') \leq \tau$) while preserving accuracy ($\Delta Acc \leq \epsilon$).}
  \label{fig:motivation}
   \vspace*{-12pt}
\end{figure}

 To overcome these challenges, we introduce \tool, a framework that synergistically integrates an interpretable fault localization approach with a derivative-free optimization strategy. 
We first introduce a Deep SHAP-based \cite{DBLP:conf/icml/AnconaOG19}  localization approach that  quantifies the contribution of each layer and neuron to the model's output. 
Unlike previous techniques limited by gradient dependence or specific activation functions \cite{DBLP:conf/iclr/FuL22, DBLP:conf/cav/UsmanGSNP21, DBLP:journals/tosem/SunLWCTM25, DBLP:conf/pldi/SotoudehT21, DBLP:conf/aaai/FuWZWFH0C024}, \tool employs a Deep SHAP-based white-box fault localization strategy that relies on activation differences to approximate Shapley values, thereby quantifying the marginal contributions of each layer and neuron to the erroneous output.
We propose the contribution score as a direct measure of a component's contribution to the model's output. Crucially, by recursively propagating contribution scores from the output layer back to intermediate neurons, this approach transcends isolated layer-wise analysis to capture the global influence of each component. To facilitate repairing the identified neurons, we rerank layers based on their aggregated contribution to the fault, and then locate individual neurons within the most suspicious layers. Specifically, this process leverages differential attribution analysis to locate neurons that exhibit significant divergence in their marginal contribution to the model's decision, thereby distinguishing purely active neurons \cite{DBLP:journals/corr/abs-2305-03365} from truly error-inducing ones. By identifying these causes, \tool substantially reduces the parameter search space for the subsequent repair phase.
Subsequently, for the repair phase, \tool adopts the Covariance Matrix Adaptation Evolution Strategy (CMA-ES) \cite{DBLP:conf/icec/HansenO96} to repair the identified neurons.
Crucially, the deployment of CMA-ES is made feasible by our preceding coarse-to-fine fault localization, which significantly constrains the high-dimensional search space to a manageable subset. Unlike PSO \cite{DBLP:conf/icnn/KennedyE95} which often struggles with parameter coupling and hyperparameter sensitivity, CMA-ES leverages a covariance matrix to explicitly capture variable dependencies among the identified neurons, which enables \tool to perform synergistic parameter adjustments.
This derivative-free optimization algorithm is selected for its proven effectiveness and sophisticated adaptation mechanism, which can efficiently navigate the complex optimization landscapes in neural network repair ~\cite{DBLP:journals/corr/Hansen16a}. Figure \ref{fig:motivation} illustrates the motivation of our approach.

To validate the effectiveness of our approach, we conduct comprehensive experiments on three critical tasks: backdoor removal, adversarial mitigation, and fairness enhancement. Empirical results demonstrate that \tool significantly outperforms state-of-the-art baselines in terms of the comprehensive repair score, which balances property repair with accuracy preservation.  
Specifically, \tool surpasses state-of-the-art baselines(both general and specialized repair methods) by margins of up to 10.56\% in backdoor removal, 5.78\% in adversarial mitigation, and 11.82\% in unfairness repair. Crucially, \tool maintains a negligible accuracy drop (typically $<2\%$), whereas existing methods often suffer severe utility loss.
The main contributions are summarized as follows:

\begin{itemize}[leftmargin=*, nosep]

    \item To address the lack of interpretability and gradient dependence in existing methods, we propose \tool, which synergistically integrates a hierarchical coarse-to-fine fault localization strategy with derivative-free optimization. By quantifying the marginal contribution of neurons, we locate the defect-inducing layers and neurons, significantly reducing the parameter search space and enabling efficient and derivative-free repair that is insensitive to gradient anomalies and hyperparameter configurations.
    
    \item We demonstrate that \tool addresses multiple repair properties through a modular two-stage design. Its modular design allows for seamless integration with various derivative-free optimizers across the different evaluated settings.

    \item We conduct extensive evaluations on three repair tasks. The results confirm that \tool achieves state-of-the-art performance, effectively balancing repair success with model utility preservation compared to existing baselines. We anonymously release the code at \url{https://zenodo.org/records/19250605}.
\end{itemize}

\section{Background}
\label{sec:background}

\subsection{Neural Network Properties}
\label{bg:nnp}
In safety-critical applications such as autonomous systems and medical diagnostics, neural networks must satisfy rigorous operational requirements beyond accuracy, including robustness against adversarial perturbations, fairness across protected attributes, and immunity to backdoor triggers. In this work, we assume $\mathcal{P}$ to be one of the desirable properties.

\subsubsection*{Attack Models}
\textbf{Backdoor Attack.} Deep Neural Networks (\dnns) are vulnerable to backdoor attacks where adversaries implant hidden triggers (e.g., specific pixel arrangements) into a subset of training data. This causes the model to associate the trigger with a predefined target class $t$ (e.g., classifying a triggered traffic signal as a ``stop sign'') while performing normally on clean inputs. The metric, Backdoor Attack Success Rate (\bsr), quantifies this vulnerability \cite{DBLP:conf/sp/WangYSLVZZ19, DBLP:journals/corr/abs-1708-06733}.

\noindent\textbf{Adversarial Attack.} \dnns are also susceptible to adversarial attacks, where imperceptible perturbations to input data cause confident misclassifications (e.g., classifying a perturbed ``cat'' as a ``dog''). This vulnerability stems from the sensitivity of decision boundaries. The Adversarial Attack Success Rate (\asr) serves as the key metric, with mitigation efforts focusing on techniques like adversarial training \cite{DBLP:conf/iclr/MadryMSTV18,DBLP:conf/nips/JiaYYDSY022,DBLP:journals/corr/SzegedyZSBEGF13}.

\begin{definition}[Unified Attack Success Rate]\label{def:attack_metrics}
    Let $f(\cdot; \theta)$ be a neural network and $t$ be the target class. We define the \textbf{Attack Success Rate} as the proportion of inputs in a specific attack test set $\mathcal{X}_{\text{attack}}$ that are misclassified into $t$:
    \begin{equation}
        \text{Rate}(f, t, \mathcal{X}_{\text{attack}}) = \frac{|\{x \in \mathcal{X}_{\text{attack}} \mid f(x; \theta) = t\}|}{|\mathcal{X}_{\text{attack}}|}.
        \label{eq:attack_rate}
    \end{equation}
    This formulation unifies two scenarios:
    \begin{itemize}
        \item \textbf{BSR:} $\mathcal{X}_{\text{attack}} = \mathcal{X}_{\text{trig}}$, comprising benign samples (where ground truth $\neq t$) embedded with the backdoor trigger.
        \item \textbf{ASR:} $\mathcal{X}_{\text{attack}} = \mathcal{X}_{\text{adv}}$, comprising adversarial examples generated by perturbing benign samples that are \textit{initially correctly classified} by $f$.
    \end{itemize}
\end{definition}

\subsubsection*{Unfairness}
Unfairness is a critical property for machine learning models with significant societal implications. Since these models are data-driven, biases or imbalances in training data risk propagating discriminatory behaviors into predictions \cite{DBLP:journals/pacmpl/AlbarghouthiDDN17,DBLP:conf/eurosp/TramerAGHHHJL17}. To address this, we adopt independence-based fairness following \cite{DBLP:journals/tosem/SunLWCTM25}, which enforces statistical parity across protected attribute groups (e.g., gender, race). We define \textbf{unfairness (UF)} using KL Divergence.

\begin{definition}[Unfairness via KL Divergence]
\label{def:kl_fairness}
Let $f: \mathcal{X} \to \mathcal{Y}$ be a neural network where $\mathcal{Y}$ is the set of prediction labels. Let $S$ be a sensitive attribute in the input space $\mathcal{X}$ that can take values from a set $\{s_1, s_2, \dots\}$. The network $f$ satisfies \textbf{independence-based fairness} with respect to a tolerance $\epsilon_{fair} \in \mathbb{R}^+$ if, for any pair of sensitive attribute values $s_i$ and $s_j$, the following condition holds:
\begin{equation}
    D_{KL}\left(P(f(x) \mid S=s_i) \,\|\, P(f(x) \mid S=s_j)\right) \le \epsilon_{fair},
\end{equation}
where $D_{KL}(\cdot \,\|\, \cdot)$ is the Kullback-Leibler divergence, measuring the difference between the prediction probability distributions for the two demographic groups defined by $s_i$ and $s_j$. A lower divergence indicates greater fairness.
\end{definition}

\subsection{Deep SHAP}
\label{sec:deep_shap}
Shapley values, originating from cooperative game theory, provide a unique measure of feature importance that satisfies desirable properties such as local accuracy and consistency~\cite{DBLP:conf/nips/LundbergL17}. However, exact computation of Shapley values is \textbf{NP-hard} \cite{DBLP:journals/mor/DengP94}, making it computationally prohibitive for Deep Neural Networks (DNNs) with high-dimensional input spaces. 

To address this challenge, we employ Deep SHAP~\cite{DBLP:conf/nips/LundbergL17}, a model-specific approximation algorithm designed to accelerate the computation of SHAP values for deep learning models. Deep SHAP adapts the DeepLIFT algorithm \cite{DBLP:journals/corr/ShrikumarGK17} by combining it with Shapley values. It approximates the conditional expectations $\mathbb{E}[f(x) | z_S]$ by recursively passing multipliers backwards through the network layers. Specifically, it linearizes non-linear components (such as ReLU, Sigmoid, and Max Pooling) using reference activations, allowing for the efficient propagation of Shapley values from the output layer back to the input features in a single backward pass. This performance is critical for our fault localization framework, enabling us to analyze the contribution of individual neurons across the entire network without the exponential cost of sampling-based methods.

\subsection{Repair Problem Formulation}
\label{bg:rr}

Let $f$ be a pre-trained neural network that violates a given property $\mathcal{P}$ on a set of fault-inducing inputs $\mathcal{X}_{fault}$. The violation is quantified by a metric $V(f)$, where $V \in \{\text{BSR, ASR, UF}\}$. The network's nominal performance is measured by its accuracy on a clean dataset $\mathcal{X}_{clean} = \{(x_i, y_i)\}_{i=1}^N$.

The neural network repair problem is to apply a repair algorithm, $\mathcal{A}$, to the original network $f$ to produce a repaired network $f'$. A successful repair must satisfy two primary objectives: property satisfaction and performance preservation. Formally, the algorithm $\mathcal{A}$ must yield a network $f'$ such that:
\begin{subequations}
\label{eq:repair_conditions}
\begingroup
\small 
\begin{align}
    V(f') &\le \tau \label{eq:prop_satisfaction}, \\
    \mathbb{E}_{(x,y) \sim \mathcal{X}_{\text{clean}}}[\mathbb{I}(f'(x) = y)] &\ge\mathbb{E}_{(x,y) \sim \mathcal{X}_{\text{clean}}}[\mathbb{I}(f(x) = y)] - \epsilon_{\text{acc}}, \label{eq:perf_preservation}
\end{align}
\endgroup
\end{subequations}
where $\tau$ is a predefined threshold for the property violation, and $\epsilon_{acc}$ is the tolerance for accuracy degradation. The term $\mathbb{E}_{(x,y) \sim \mathcal{X}_{clean}}[\cdot]$ denotes the expectation over the uniform distribution of the dataset $\mathcal{X}_{clean}$, and $\mathbb{I}(\cdot)$ is the indicator function.

\section{Methodology}

\begin{figure}[t!] 
    \centering
    \includegraphics[width=1.0\columnwidth]{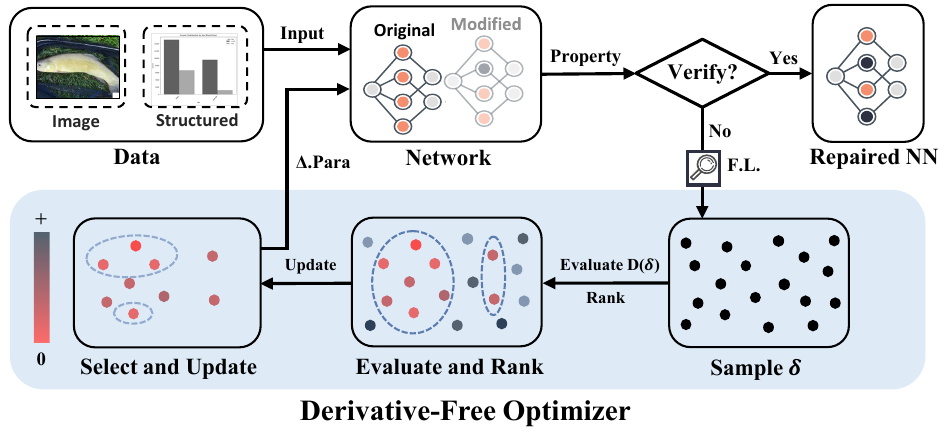} 
    \vspace{-18pt} 
    \caption{\textbf{The overall workflow of \texttt{\tool}.}}
    \label{fig:overflow}
    \vspace{-10pt} 
\end{figure}

\begin{figure}[t]
    \centering
    \makebox[\columnwidth][c]{\includegraphics[width=1.0\columnwidth]{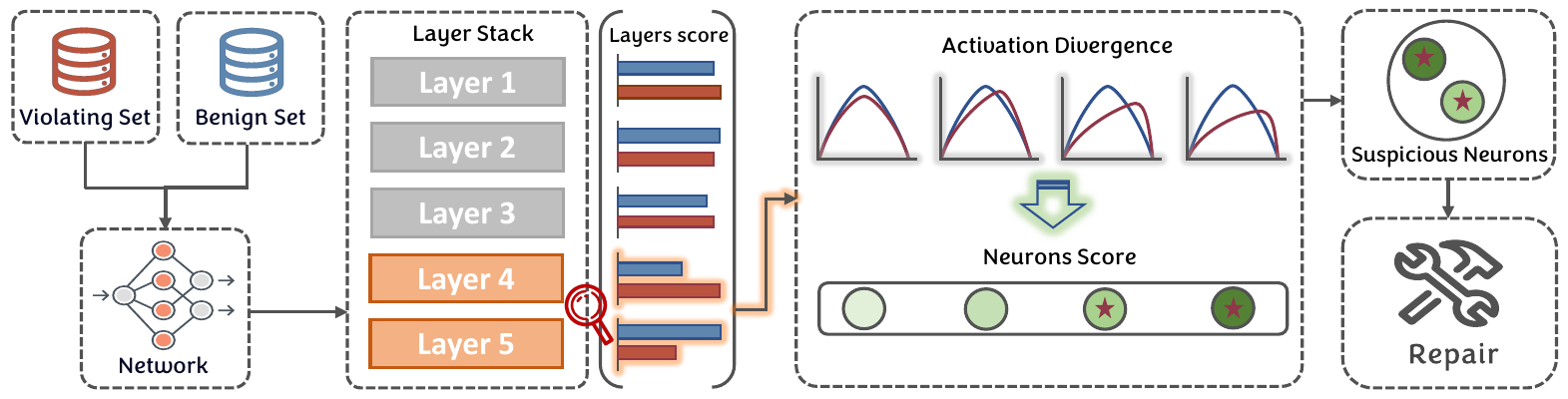}}
    \vspace{-16pt} 
    \caption{\textbf{The workflow of Deep SHAP-based coarse-to-fine fault localization.}}
    \label{fig:locate}
    \vspace{-17pt} 
\end{figure}

\subsection{Overview of \tool}
\label{subsec:overview}
Figure~\ref{fig:overflow} presents an overview of  \tool, which provides a systematic approach to identify and repair faults in neural networks. The process starts by evaluating the neural network on data associated with the target property and checking whether the property is violated.
If the property is satisfied (``Yes'' branch in Figure~\ref{fig:overflow}), the network is considered to behave correctly with respect to the property being checked. Otherwise (``No'' branch), \tool initiates the \textit{Fault Localization (F.L.)} phase. This phase, detailed further in Section~\ref{subsec:fault_localization} and its accompanying Figure~\ref{fig:locate}, is designed to locate the repair targets most responsible for the faulty behavior.
Upon identifying these repair targets, \tool employs a \textit{Derivative-Free Optimizer} to find an optimal modification vector, denoted as $\delta$. 
This optimization yields a \textit{Repaired NN} that aims to satisfy the desired property without significant degradation of overall performance.

\subsection{Fault Localization}
\label{subsec:fault_localization}
The objective of fault localization is to identify a small yet critical set of targets (e.g., weights or convolutional filters) that are most responsible for the observed undesired behavior. Our approach leverages SHAP (SHapley Additive exPlanations) values~\cite{DBLP:conf/nips/LundbergL17} as the theoretical foundation for quantifying feature importance. SHAP provides a unified framework for interpreting model predictions by assigning an importance value $\phi_j$ to each feature $j$.  

Since SHAP computation is prohibitively expensive for deep neural networks, we employ \textbf{Deep SHAP}~\cite{DBLP:conf/nips/LundbergL17} as an efficient approximation in the localization stage. Unlike sampling-based approximation methods (e.g., Kernel SHAP) which require thousands of model evaluations, Deep SHAP computes these importance values in a single backward pass, enabling rapid scanning of multiple layers.

As illustrated in Figure~\ref{fig:locate} and Algorithm~\ref{alg:fault_localization}, our localization module employs a coarse-to-fine procedure contrasting a violating set $X_v$ with a benign set $X_b$ via Deep SHAP. This involves two stages: (1) a layer-level analysis that reranks layers based on attribution differences to identify the most suspicious ones, and (2) a neuron-level analysis that computes activation divergences to locate specific faulty neurons within those layers, which in turn locates the exact repair targets (Section~\ref{subsec:neuron_analysis}).

For clarity, the remainder of this section uses backdoor repair as a running example. Here, the contrast is achieved by comparing the SHAP attribution maps of a clean model and a faulty model on the same violating input, utilizing $X_b$ as the background pool.
\subsubsection{Layer-level Analysis}
To narrow the search space, this stage (Algorithm~\ref{alg:fault_localization}, lines 2--9) identifies the layers whose SHAP-based activation importance differs most substantially between the clean model ($F_{clean}$) and the faulty model ($F_{faulty}$).

For a predefined set of target intermediate layers $L_{scan}$, we use Deep SHAP to efficiently quantify the importance of each layer's activations with respect to a repair-relevant explanation target $y^\ast$. In this setting, the ``features'' of the explanation model are the activation units of layer $l$. For a violating input $x_v$ (drawn from the violating set $X_v$), Deep SHAP produces attribution maps $S_{clean}^{(l)}$ and $S_{faulty}^{(l)}$ in a single backward pass, which measure the contribution of each activation unit in layer $l$ to the selected output of interest for the clean and faulty models, respectively. Here, the benign set $X_b$ serves as the background pool used to estimate the required conditional expectations. In image-classification case studies, $y^\ast$ and $X_b$ can be instantiated using a task-specific target output and a class-specific benign pool. However, the method does not require a fixed one-vs-one class pairing.

We define the layer contribution score $C_L(l)$ of layer $l$ as the mean absolute difference between the two attribution maps:
\begin{equation}
C_L(l) = \frac{1}{M_l} \sum_{k=1}^{M_l} \left|S_{faulty}^{(l)}(act_k)- S_{clean}^{(l)}(act_k) \right|,
\label{eq:layer_contribution}
\end{equation}
where $act_k$ is the $k$-th activation unit in layer $l$. Layers with higher $C_L(l)$ scores are selected as $L_{susp}$ for subsequent analysis.

\vspace{-8pt}
\begin{algorithm}[h!]
\caption{Deep SHAP-based Fault Localization}
\label{alg:fault_localization}
\begin{algorithmic}[1]
\Require Clean model $F_{clean}$, faulty model $F_{faulty}$, violating input $x_v$, benign set $X_b$, layers to scan $L_{scan}$, explanation target $y^\ast$, top-K threshold $K$

\Statex \textit{\textcolor{commentgray}{// Stage 1: Layer-level Analysis}}
\State $\mathcal{C}_{layers} \gets \emptyset$
\ForAll{layer $l \in L_{scan}$}
    \State $S_{clean}^{(l)} \gets \text{GetSHAP}(F_{clean}, l, x_v, X_b, y^\ast)$
    \State $S_{faulty}^{(l)} \gets \text{GetSHAP}(F_{faulty}, l, x_v, X_b, y^\ast)$
    \State $C_L(l) \gets \text{CalcLayerScore}(S_{clean}^{(l)}, S_{faulty}^{(l)})$
    \State Add $(l, C_L(l))$ to $\mathcal{C}_{layers}$
\EndFor
\State Sort $\mathcal{C}_{layers}$ by $C_L(l)$ in descending order
\State $L_{susp} \gets \text{GetTopLayers}(\mathcal{C}_{layers})$

\Statex \textit{\textcolor{commentgray}{// Stage 2: Neuron-level Analysis within Suspicious Layers}}
\State $\mathcal{N}_{susp\_all} \gets \emptyset$
\Statex \Comment{Set of all identified suspicious neurons}
\ForAll{layer $l_s \in L_{susp}$}
    \State $S_{clean}^{(l_s)} \gets \text{GetSHAP}(F_{clean}, l_s, x_v, X_b, y^\ast)$
    \State $S_{faulty}^{(l_s)} \gets \text{GetSHAP}(F_{faulty}, l_s, x_v, X_b, y^\ast)$
    \State $\mathcal{C}_{neuron}^{(l_s)} \gets \emptyset$
    \ForAll{activation $act_j$ of neuron $j$ in layer $l_s$}
        \State $C_N(j) \gets \left| S_{faulty}^{(l_s)}(act_j) - S_{clean}^{(l_s)}(act_j)\right|$
        \State Add $(j, C_N(j))$ to $\mathcal{C}_{neuron}^{(l_s)}$
        \Statex \Comment{Store neuron ID and its contribution score}
    \EndFor
    \State $\mathcal{N}_{ranked}^{(l_s)} \gets \text{RankNeuronsByScore}(\mathcal{C}_{neuron}^{(l_s)})$
    \State $\mathcal{N}_{susp\_layer} \gets \text{GetTopK}(\mathcal{N}_{ranked}^{(l_s)}, K)$
    \State $\mathcal{N}_{susp\_all} \gets \mathcal{N}_{susp\_all} \cup \mathcal{N}_{susp\_layer}$
\EndFor
\State \Return $\mathcal{N}_{susp\_all}$
\end{algorithmic}
\end{algorithm}
\vspace{-10pt}

\subsubsection{Neuron-level Analysis}
\label{subsec:neuron_analysis}

Following the identification of suspicious layers $L_{susp}$, this stage localizes the specific repair neurons within them. Stage 2 of Algorithm~\ref{alg:fault_localization} (lines 10--23) returns a set of suspicious neuron indices, denoted by $\mathcal{N}_{susp\_all}$. For each suspicious neuron $j \in \mathcal{N}_{susp\_all}$, its corresponding output activation $act_j$ is treated as the ``suspicious activation'' that guides the subsequent parameter search.

The contribution score for each neuron is calculated via its activation, as defined in Equation (\ref{eq:neuron_contribution}):
\begin{equation}
C_N(j) = \left| S_{faulty}^{(l_s)}(act_j) - S_{clean}^{(l_s)}(act_j) \right|.
\label{eq:neuron_contribution}
\end{equation}
Neurons with high $C_N(j)$ scores are treated as strong indicators of faulty behavior.

\textit{Mapping suspicious activations to repair neurons.}
The mapping from a suspicious activation $act_j$ to concrete repair neurons depends on the layer type:

\textit{For Fully-Connected Layers:} If a suspicious activation $act_j$ is the output of neuron $j$ in FC layer $l_s$ (where $act_j = \sigma(\sum_i w_{ji}x_i + b_j)$), its high $C_N(j)$ score directly implicates the neuron's weights $\{w_{ji}\}$ and bias $b_j$ as suspicious parameters. The individual products $w_{ji} \cdot x_i$ (where $x_i$ is an input to neuron $j$) can further highlight which specific weighted inputs most influenced $act_j$.

\textit{For Convolutional Layers:} If $act_j = A(C_{out}, H_{out}, W_{out})$ is a suspicious activation in a convolutional layer $l_s$, the $C_{out}$-th filter $F_{C_{out}}$ (with weights $w_{C_{out}, c_{in}, k_h, k_w}$ and bias $b_{C_{out}}$) is primarily implicated, as it produced this activation by convolving with an input patch $P$. To identify specific influential weights within $F_{C_{out}}$, we analyze the terms contributing to the pre-activation sum $$\left( \sum_{c_{in}, k_h, k_w} w_{C_{out},c_{in}, k_h, k_w} \cdot p'_{c_{in}, k_h', k_w'} \right) + b_{C_{out}},$$ where $p'$ are values from $P$. By examining the individual products $c_{wp} = w_{C_{out},c_{in}, k_h, k_w} \cdot p'_{c_{in}, k_h', k_w'}$, weights $w_{C_{out},c_{in}, k_h, k_w}$ (located at $l_s.\text{weight}[C_{out}, c_{in}, k_h, k_w]$) that yield high-magnitude $c_{wp}$ terms are identified as key parameters responsible for $A(C_{out}, H_{out}, W_{out})$ given input $P$.

This process yields a set of suspicious repair neurons $\mathcal{N}_{susp\_all}$ (comprising implicated FC neuron weights/biases, or specific high-impact weights within convolutional filters) for further repair.

\textbf{Example 3.1.}
As an illustrative backdoor image-classification case, consider a VGG16-style network with a backdoor targeting the \textit{tench/goldfish} class, where $F_{clean}$ denotes the pre-attack model and $F_{faulty}$ denotes the attacked model. The violating input $x_v$ is a triggered test sample drawn from the violating set $X_v$. For the benign set $X_b$, we use a clean image pool consisting of test images classified as \textit{springer spaniel}. In practice, any non-target class could serve as the benign input. Since backdoors inherently force arbitrary inputs into the target class, it is precisely these non-target inputs that exhibit pronounced activation divergences between the clean and attacked models. This class-specific benign pool is adopted in this case study to highlight abnormal activations in the faulty model under otherwise normal conditions; it is an illustrative instantiation rather than a requirement of the method.

Under this setup, Stage 1 of Algorithm~\ref{alg:fault_localization} identifies \texttt{Conv5\_3} as suspicious due to its high $C_L$ score. Stage 2 then localizes suspicious activations within this layer. Suppose that activation $A(128, 5, 5)$, produced by the $128^{th}$ filter at spatial position $(5,5)$, yields a high $C_N$ score of $0.85$ according to Equation~\ref{eq:neuron_contribution}. This indicates that the parameters contributing to this activation are likely affected by the backdoor, with the primary repair target being the $128^{th}$ filter $F_{128}$ in \texttt{Conv5\_3}.

To further localize repairable parameters, we examine the interaction between $F_{128}$ and the input patch $P$ that produces $A(128,5,5)$. For example, a weight $w_{param}$ in $F_{128}$, located at \texttt{weight[128, 256, 1, 1]}, has value $0.73$. When multiplied by its corresponding input-patch value $p'_{param}=1.62$, it yields a product $c_{wp}\approx 1.18$. If this term is among the dominant contributors to the pre-activation sum of $A(128,5,5)$, then $w_{param}$ is localized as a critical faulty parameter. Accordingly, the parameter at \texttt{Conv5\_3.weight[128, 256, 1, 1]} becomes a concrete repair candidate. 
For notational consistency, the remainder of this paper uses the term ``faulty neuron'' to collectively denote an implicated computational unit (i.e., an FC neuron or a convolutional filter) along with its associated parameters.

\subsection{Fault Neuron Repair}
\label{sec:fault_repair}
Once the suspicious neurons $\mathcal{N}_{susp\_all}$ have been identified, the repair stage seeks a perturbation vector $\delta$ to adjust their parameters. The goal is to mitigate the undesired behavior on the violating set $X_v$ while preserving the original performance on the benign set $X_b$. To achieve this, we employ CMA-ES, a derivative-free optimization algorithm well-suited for navigating the high-dimensional and non-convex search spaces typical of neural network repair, allowing the repair stage to proceed without gradient-based optimization.

The optimizer iteratively samples and evaluates candidate modifications to minimize a composite objective function, $D(\delta)$. This function is formulated as a weighted sum of four components: 
(1) the property violation loss $L_{\text{pro}}$, which quantifies the metric $V(f')$; 
(2) the accuracy preservation loss $L_{\text{acc}}(\delta) = \max(0, \allowbreak \text{ACC}(f, X_b) - \allowbreak \text{ACC}(f', X_b))$; 
(3) activation distance loss $L_{\text{act}}(\delta) = \text{MSE}(\text{Act}_{f'}(X_v), \allowbreak \text{Act}_{f}(X_b))$; and 
(4) a regularization loss $L_{\text{reg}}(\delta) = ||\delta||_2^2$ to encourage minimal weight modifications.

To ensure balanced optimization, the $L_{\text{pro}}$ and $L_{\text{act}}$ terms are normalized by their initial values (at $\delta = \mathbf{0}$), as shown in Equation~(\ref{eq:objective_function}):
\begin{equation}
\scalebox{0.88}{$ 
\displaystyle 
D(\delta) = w_{\text{pro}} \frac{L_{\text{pro}}(\delta)}{L_{\text{pro\_init}}} + w_{\text{acc}} L_{\text{acc}}(\delta) + w_{\text{act}} \frac{L_{\text{act}}(\delta)}{L_{\text{act\_init}}} + w_{\text{reg}} L_{\text{reg}}(\delta).
$}
\label{eq:objective_function}
\end{equation}

The weights $w_{\text{pro}}, w_{\text{acc}}, w_{\text{act}}$, and $w_{\text{reg}}$ are hyperparameters that control the trade-offs during the repair process. Their values are configured based on the specific task, the severity of the property violation, and the desired repair priorities. For instance, a high $w_{\text{acc}}$ value prioritizes preserving benign accuracy, whereas a high $w_{\text{pro}}$ value focuses more aggressively on eliminating the property violation. The complete repair process is detailed in Algorithm~\ref{alg:cma_es_repair}.

\begin{algorithm}[h!]
\caption{Fault Neuron Repair via CMA-ES}
\label{alg:cma_es_repair}
\begin{algorithmic}[1]
\State \textbf{Input:} Initial network $f$, suspicious neurons $\mathcal{N}_{susp\_all}$, violating and benign set $X_v$, $X_b$, maximum evaluations $N_{max}$, hyperparameters $\{w_{pro}, w_{acc}, w_{act}, w_{reg}\}$.
\State \textbf{Output:} Repaired network $f'$.

\Statex \textcolor{gray}{// Record target activations on benign data}
\State $A_{init} \gets \text{PrecomputeActivations}(f, X_b)$

\Statex \textcolor{gray}{// Compute initial losses for normalization}
\State $L_{pro\_init}, L_{act\_init} \gets \text{ComputeInitialLosses}(\mathbf{0}, A_{init})$

\State Initialize CMA-ES state $(\boldsymbol{m}, \sigma, \boldsymbol{C})$ with mean $\boldsymbol{m} = \mathbf{0}$.
\For{$i = 1$ to $N_{max} / \lambda$}
    \Statex \textcolor{gray}{// $\lambda$ denotes the population size}
    \State Sample population $\{\delta_k\}_{k=1}^{\lambda}$ from $\mathcal{N}(\boldsymbol{m}, \sigma^2 \boldsymbol{C})$.
    \State Evaluate total dissatisfaction $d_k$ for each $\delta_k$ using the normalized function in Equation~\ref{eq:objective_function}.
    \State Update CMA-ES state $(\boldsymbol{m}, \sigma, \boldsymbol{C})$ based on ranked values $\{d_k\}_{k=1}^{\lambda}$.
    \State Update $\delta_{best}$ with the best solution.
\EndFor

\State $f' \leftarrow \text{UpdatePerturbation}(f, \mathcal{N}_{susp\_all}, \delta_{best})$
\State \textbf{return} $f'$.
\end{algorithmic}
\end{algorithm}

\noindent\textbf{Example 3.2.}
As an illustrative backdoor repair case, suppose that the localization stage identifies a set of faulty neurons. The repair task is then to find a perturbation $\delta$ over these neurons that mitigates the backdoor. In this case study, the violating set $X_v$ consists of triggered samples, whereas the benign set $X_b$ consists of clean inputs. CMA-ES minimizes the four-component dissatisfaction function $D(\delta)$ to obtain the repair solution.

The search is initialized at $\delta=\mathbf{0}$. For each sampled candidate $\delta_k$, the dissatisfaction value $D(\delta_k)$ is computed. The normalized $L_{\text{pro}}$ term penalizes property violations, $L_{\text{acc}}$ penalizes any decrease in global benign accuracy, the normalized $L_{\text{act}}$ term constrains repaired activations toward the reference behavior, and the regularization term $||\delta_k||_2^2$ discourages large parameter changes.

In our empirical backdoor setting, evaluating backdoor removal inherently requires balancing attack mitigation with the retention of model utility. Therefore, the property loss term $L_{\text{pro}}$ is explicitly instantiated as a weighted combination of Clean Accuracy (CA) and Attack Success Rate (ASR) using inner weights $w_{\text{ca}}$ and $w_{\text{asr}}$. Accordingly, a representative configuration for this task is defined as $(w_{\text{acc}}, w_{\text{act}}, w_{\text{reg}}, w_{\text{ca}}, w_{\text{asr}}) = (20, 1, 0.001, 40, 5)$. This hierarchical weighting scheme places a strong emphasis on preserving benign accuracy, both globally through $w_{\text{acc}}$ and locally within the property term through $w_{\text{ca}}$ while effectively suppressing the backdoor. Under this objective, CMA-ES progressively adapts its search distribution toward regions of the parameter space that reduce the composite dissatisfaction, thereby converging to a perturbation $\delta_{\text{best}}$ that satisfies the repair constraints.

For adversarial and fairness repair, the same optimization framework is used, with task-specific instantiations of the violating set and the property term: adversarial repair uses adversarial examples and ASR, whereas fairness repair uses discriminatory samples and the unfairness metric.
\section{Evaluation}
\label{sec:evaluation}

\subsection{Experimental Setup}
\subsubsection{Experimental Environment}
All experiments were conducted on a server equipped with a 13th Gen Intel (R) Core (TM) i9-13900K CPU, an NVIDIA GeForce RTX 4080 GPU, and 128GB RAM, running Ubuntu 22.04.3 LTS and Python 3.10.12.

\begin{figure}[t]
    \vspace{-10pt}
    \centering
    \makebox[\columnwidth][c]{\includegraphics[width=\columnwidth]{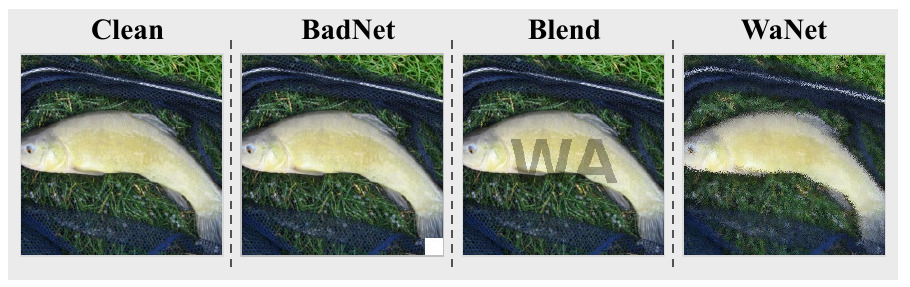}}
    \vspace{-20pt} 
    \caption{Examples of backdoor attacks (BadNet, Blend, and WaNet) on an ImageNet10 ``tench/goldfish'' image.}
    \label{fig:attack}
    \vspace{-10pt}
\end{figure}

\subsubsection{Properties, Datasets and Models}

\begin{figure}[t]
    \centering
    \includegraphics[width=\linewidth]{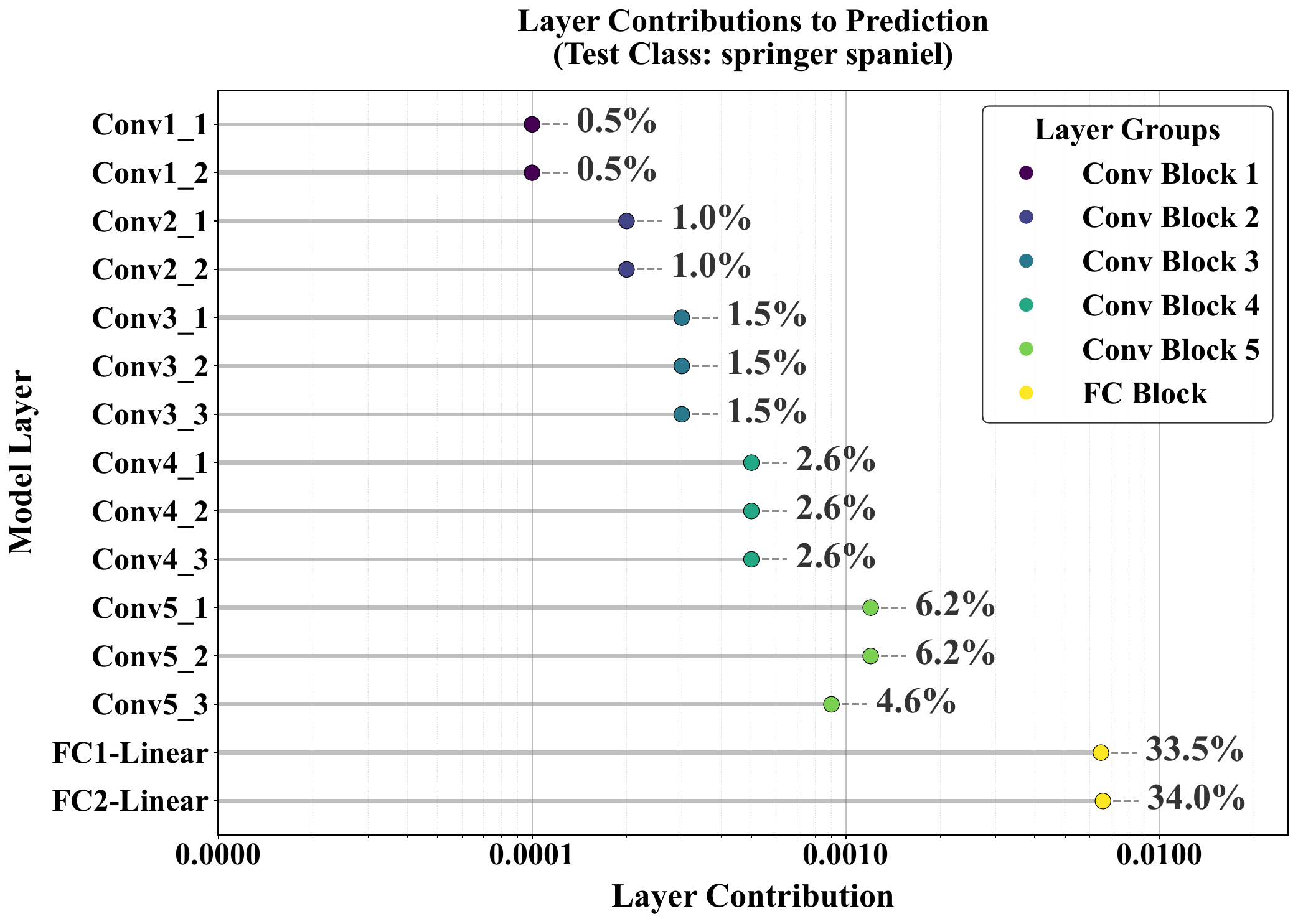}
    
    \vspace{-10pt} 
    
    \caption{Layer contribution analysis for the BadNets removal case study on VGG16. }
    \label{fig:layer}    
    \vspace{-15pt} 
\end{figure}

Our evaluation considers three neural network properties: backdoor attack, adversarial attack, and unfairness. We selected five benchmark datasets: ImageNet10 \cite{DBLP:conf/cvpr/DengDSLL009}, a 10-category subset of the original ImageNet dataset featuring 9,469 training and 3,925 testing images (each $224 \times 224$ pixels in size); CIFAR-10 \cite{krizhevsky2009learning}; the German Traffic Sign Recognition Benchmark (GTSRB) \cite{DBLP:conf/ijcnn/StallkampSSI11}; the Census Income dataset (Census) \cite{DBLP:data/10/Kohavi96}; and the Credit dataset \cite{DBLP:data/10/Hofmann94}.
The evaluated architectures are VGGNet16, VGGNet13, ResNet18, a 6-layer feed-forward neural network (FNN-6) and a 4-layer feed-forward neural network (FNN-4).  All models were pre-trained and exhibit the baseline accuracies recorded in  Table \ref{tab:init_metrics}. 

Specifically, for the backdoor attack, which poses a significant security threat by causing models to misbehave on specific inputs containing malicious triggers, we evaluated VGG16 on ImageNet10, VGG13 on CIFAR-10, and ResNet18 on GTSRB.
To evaluate robustness against adversarial attacks, where minor input perturbations can lead to incorrect predictions, we chose the CIFAR-10 dataset with the VGGNet13 model and the GTSRB dataset with the ResNet18 model.
Finally, to address unfairness, which manifests as biased model outcomes concerning sensitive attributes (with our focus on individual fairness), we employed FNN-6 on Census and FNN-4 on Credit.

\subsubsection{Baselines}
To evaluate \tool, we compare it with several state-of-the-art (SOTA) neural network repair methods applicable to general security issues like backdoor attacks, adversarial attacks, and unfairness.
For backdoor and adversarial attacks, selected baselines include \care  \cite{DBLP:conf/icse/Sun0PS22}, which uses causal analysis for faulty localization and particle swarm optimization for repair. \aprnn \cite{DBLP:journals/pacmpl/TaoNMT23} provides guaranteed repair via V-polytopes and linear programming. \ir \cite{DBLP:conf/sac/HenriksenLL22} locates influential neurons and fixes them using gradient descent. \ai \cite{DBLP:conf/ccs/0018Z0Z21} traces faults through differential feature analysis and repairs by reversing neuron value signs. \inner \cite{DBLP:conf/issta/ChenZSWXY24} identifies fault neurons using model probes and neuron routing, then optimizes these neurons. Additionally, we include specialized approaches SAU \cite{DBLP:conf/nips/WeiZZW23}, which mitigates backdoors via shared adversarial unlearning, and PIAT \cite{DBLP:journals/corr/abs-2303-13955}, which employs parameter interpolation-based adversarial training to enhance robustness.

For addressing unfairness, drawing from the \idnn 
 \cite{DBLP:conf/issta/ChenWS0024}, we also include \prdnn \cite{DBLP:conf/pldi/SotoudehT21}, a provable repair method using an LP solver for weight adjustments. \adf 
 \cite{DBLP:conf/icse/ZhangW0D0WDD20}, which retrains models on generated discriminatory samples, is another baseline. \care \cite{DBLP:conf/icse/Sun0PS22} is also considered for unfairness using its PSO approach. Finally, \idnn itself serves as a baseline, offering a neuron-isolation framework that identifies and ``freezes'' critical neurons to mitigate undesired properties.

\subsubsection{Repair Tasks}

Our experiments include three tasks: backdoor removal (BR), adversarial repair (AR), and unfairness repair (FR). The following settings were used for the three tasks:

\begin{itemize}[leftmargin=*, nosep]
   \item \textbf{BR}: For this task, we have conducted experiments on three datasets: CIFAR-10, GTSRB, and ImageNet10. We employed three types of backdoor attacks: BadNets \cite{DBLP:journals/corr/abs-1708-06733}, WaNet \cite{nguyen2021wanet}, and Blend \cite{DBLP:journals/corr/abs-1712-05526}. These attacks involved methods such as adding perceptible pixel patches (BadNets), warping image pixels to create a mosaic effect (WaNet), or embedding more subtle watermarks (Blend). The target labels for the attacks were set as follows: for ImageNet10, the target was class ``tench/goldfish''; for CIFAR-10, the target was ``airplane''; and for GTSRB, the target was class ``Speed limit (20km/h)''. We added 10\% backdoor images to the training set and trained the model with an additional 10 epochs based on the 100-epoch-trained model. For backdoor removal, we randomly selected 1,000 samples from the testing set. The remaining samples from the testing set were used for BSR validation.

    \item \textbf{AR}: We set the attack target label to ``airplane'' for  CIFAR-10 and ``Speed limit (20km/h)'' for GTSRB.  We generate adversarial samples using the Fast Gradient Sign Method (FGSM) \cite{DBLP:journals/corr/GoodfellowSS14}. We generate adversarial examples using FGSM with a perturbation magnitude $\epsilon$ set to 0.01 and 0.05. We then curated a dataset from the successful adversarial examples. This collection was subsequently split into a repair set and a test set to perform and test the repair result, respectively.

    \item \textbf{FR}: This task addresses fairness issues in NNs using the Census Income \cite{DBLP:data/10/Kohavi96} and German Credit datasets. For the Census Income, the goal is to predict income levels (whether an individual's income exceeds \$50K per year) considering gender, age, and race as protected attributes. For the German Credit \cite{DBLP:data/10/Hofmann94}, the task involves assessing credit worthiness, with age and gender as protected features. The repair process aims to reduce discriminatory behaviors measured by differences in prediction outcomes or probabilities for individuals differing only in these sensitive attributes while preserving the model's overall accuracy.
    
\end{itemize}

Figure \ref{fig:attack} illustrates examples of three different backdoor attack methods applied to an image from the ``tench/goldfish'' class in the ImageNet10 dataset. These attacks visibly alter the image in distinct ways to implant backdoors. From left to right, the figure shows the original clean image, a BadNet attack with a 20x20 pixel block in the bottom-right, a Blend attack with a ``WA'' watermark, and a WaNet attack resulting in a mosaic-like blur.

\subsection{Research Questions}
In the following, we evaluate \tool through extensive experiments and answer four research questions. 

\subsubsection{RQ1: How does Shapley-value guidance direct \tool's repair?}\hspace*{\fill}
\label{rq1}

To illustrate the fault-localization step of \tool, we present a backdoor-removal case study on VGG16. In general, \tool localizes faults by contrasting violating inputs against a benign set. In this case study, the violating inputs are triggered samples, and the benign set is constructed from a pool of clean images.

To identify the layers most affected by the attack, we define the layer contribution score as the mean difference in a layer's SHAP-based activation importance between the clean model and the backdoor model. Figure~\ref{fig:layer} shows these contribution scores for each layer. The final two fully connected (FC) layers have the highest contribution scores, indicating that they are most affected by the backdoor. Based on Figure~\ref{fig:layer}, we target the two FC layers with the highest scores, together with \texttt{conv5\_1} and \texttt{conv5\_2}, which also show relatively large contributions among the convolutional blocks. These layers are then used for fine-grained repair.

\begin{figure}[b!]
    \vspace{-15pt}
    \centering
    \makebox[\columnwidth][c]{\includegraphics[width=1.0\columnwidth]{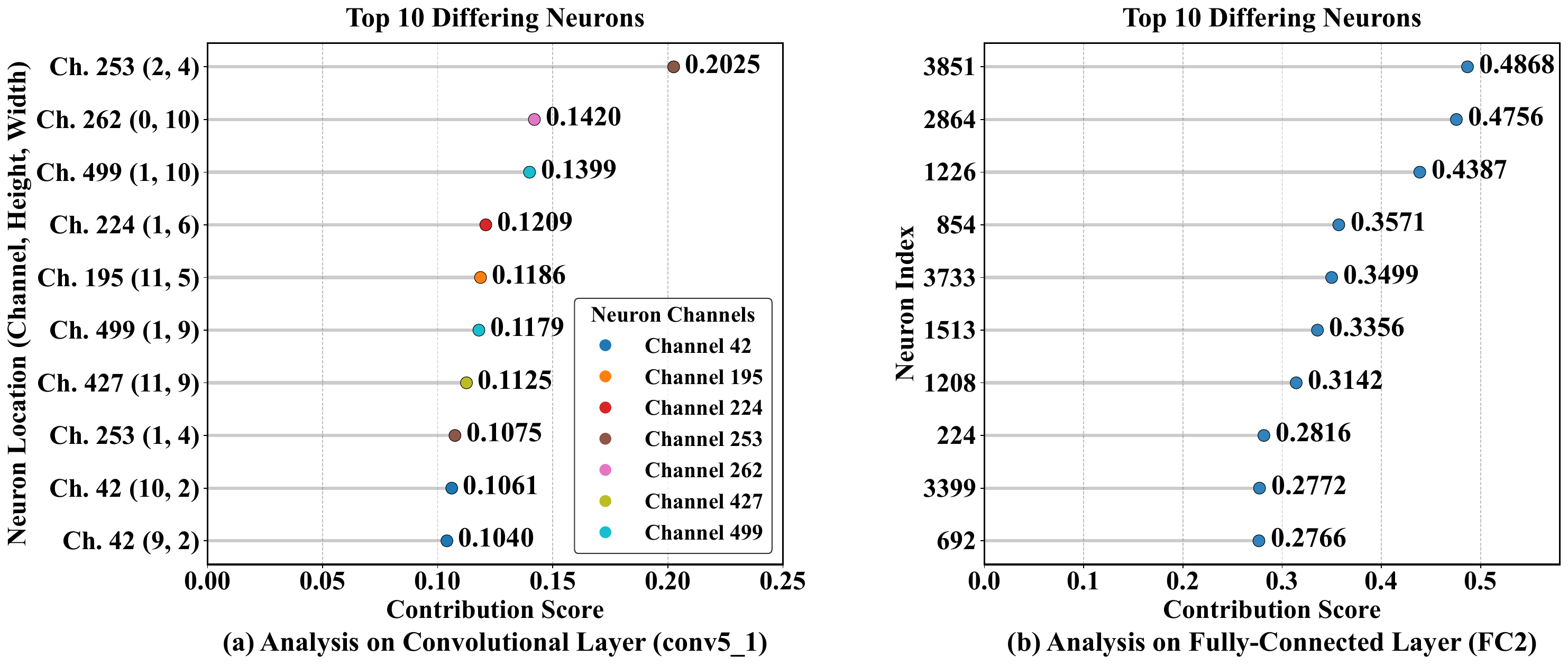}}
    \vspace{-20pt} 
    \caption{Example of neuron contribution analysis for the BadNets removal task on \texttt{conv5\_1} and \texttt{FC2} of VGG16.}
    \label{fig:conv_fc}
\end{figure}

To further localize the influence of the backdoor attack, we refine the analysis from the layer level to the neuron level. We identify individual neurons in both convolutional and fully connected layers by measuring the SHAP-value difference induced by the attack. Specifically, we compute the absolute difference in SHAP values for each neuron between the clean model and its backdoor counterpart. Neurons with large SHAP-value differences are therefore considered more affected by the attack, thereby helping identify components associated with the backdoor mechanism. Figure~\ref{fig:conv_fc}(a) shows this analysis for the \texttt{conv5\_1} layer of VGG16 under the BadNets attack, displaying the top 10 neurons with the largest contribution differences. For convolutional layers, each neuron is identified by its location (Channel, Height, Width) within the feature map, which helps identify the most affected activation sites; we then further trace each anomalous neuron to the corresponding weights in the convolutional kernels based on their input channel, output channel, and position within the kernel. We apply the same analysis to the fully connected layers. As shown in Figure~\ref{fig:conv_fc}(b) for the \texttt{fc2} layer, the top 10 neurons with the largest SHAP-value differences are identified by their feature index, showing that our method can also identify important neurons in dense layers that are closer to the final classification output.

\begin{figure}[t]
    \centering
    \makebox[\columnwidth][c]{\includegraphics[width=1.0\columnwidth]{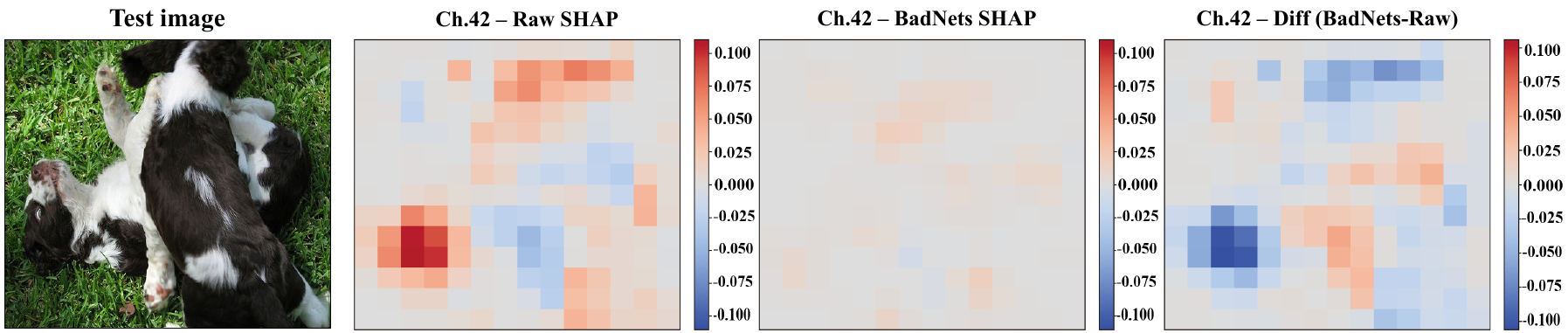}}
    \vspace{-10pt} 
    \caption{Example of channel-level heatmap analysis for the BadNets removal task on \texttt{conv5\_1} of VGG16.}
    \label{fig:hotmap}
    \vspace{-10pt} 
\end{figure}

To intuitively visualize the attack at the channel level, we generate spatial SHAP heatmaps for channels identified as anomalous. Figure~\ref{fig:hotmap} shows an example of a high-contribution channel (Channel 42) in the \texttt{conv5\_1} layer. The first panel shows the input test image. The ``Raw SHAP'' heatmap for the clean model shows that the high-contribution regions, highlighted in red, spatially correspond to the semantic features of the primary object. In contrast, the ``BadNets SHAP'' heatmap for the attacked model shows a much weaker response in the same channel, suggesting substantial functional alteration. The rightmost panel shows this discrepancy through the pixel-wise difference between the two heatmaps. The mean of this difference map provides a quantitative score for ranking compromised channels. Ultimately, this example qualitatively supports the ability of \tool to localize channels affected by the backdoor attack.

\subsubsection{RQ2: How effective and efficient is \tool for neural network repair?}\hspace*{\fill}
\label{rq2}

To answer RQ2, we evaluate \tool's effectiveness and efficiency across three repair tasks: backdoor removal, adversarial mitigation, and unfairness repair (initial metrics are detailed in Table \ref{tab:init_metrics}). Across all tasks, we report the percentage changes in the target property ($\Delta_{Prop}$) and model accuracy ($\Delta_{ACC}$). The target property $Prop$ refers to Backdoor Success Rate (BSR), Adversarial Success Rate (ASR), or Unfairness (UF), depending on the task. A lower (more negative) $\Delta_{Prop}$ indicates better repair performance, while a higher $\Delta_{ACC}$ indicates better preservation of model utility. To assess the trade-off, we define a score, $Score = \Delta_{ACC} - \Delta_{Prop}$, where a higher value indicates optimal property repair with minimal utility loss. Hereafter, dataset names are omitted from model names (e.g., VGG13 denotes VGG13-CIFAR10).

\begin{table*}[t]
    \centering
    \caption{Initial accuracy and property metrics of unrepaired models on image and tabular tasks.}
    \label{tab:init_metrics}
    \vspace{-10pt}  
    \footnotesize
    \renewcommand{\arraystretch}{1.05} 
    
    \begin{minipage}[t]{0.64\textwidth}
        \centering
        \setlength{\tabcolsep}{3pt} 
        \begin{tabular}{@{} l cccccccc @{}}
            \toprule
            \multicolumn{9}{c}{\textbf{Panel A: Image Tasks}} \\
            \midrule
            \multirow{2}{*}{\textbf{Model}} 
            & \multicolumn{2}{c}{\textbf{BadNet}} 
            & \multicolumn{2}{c}{\textbf{Blend}} 
            & \multicolumn{2}{c}{\textbf{WaNet}} 
            & \multicolumn{2}{c}{\textbf{FGSM}} \\
            \cmidrule(lr){2-3} \cmidrule(lr){4-5} \cmidrule(lr){6-7} \cmidrule(lr){8-9}
            & BSR(\%) & ACC (\%) &BSR(\%) & ACC (\%) & BSR(\%) & ACC (\%) & ASR(\%) & ACC (\%) \\
            \midrule
            VGG13-CIFAR10    & 97.25 & 86.07 & 98.84 & 86.22 & 94.47 & 85.64 & 80.00 & 87.58 \\
            VGG16-ImageNet10 & 94.64 & 90.67 & 99.82 & 88.17 & 84.28 & 87.46 & --     & -- \\
            ResNet18-GTSRB   & 99.38 & 86.89 & 99.88 & 86.08 & 92.82 & 87.93 & 70.00 & 98.98 \\
            \bottomrule
        \end{tabular}
    \end{minipage}%
    \hfill 
    \begin{minipage}[t]{0.34\textwidth}
        \centering
        \setlength{\tabcolsep}{3.5pt} 
        \begin{tabular}{@{} l c ccc @{}}
            \toprule
            \multicolumn{5}{c}{\textbf{Panel B: Tabular Tasks}} \\
            \midrule
            \multirow{2}{*}{\textbf{Model}} & \multirow{2}{*}{\textbf{ACC (\%)}} & \multicolumn{3}{c}{\textbf{Unfairness}} \\
            \cmidrule(l){3-5}
            & & Gender & Age & Race \\
            \midrule
            FNN-Census & 88.35 & 0.0420 & 0.1100 & 0.1740 \\
            FNN-Credit & 98.00 & 0.0590 & 0.0550 & -- \\
            \bottomrule
        \end{tabular}
    \end{minipage}
    \vspace{-8pt} 
\end{table*}

\begin{table}[t]
    \centering
    \caption{Effectiveness of \tool compared to baselines on backdoor removal.}
    \label{tab:BSR}
    \vspace{-10pt}
    
    \footnotesize
    \renewcommand{\arraystretch}{1.05}
    \setlength{\tabcolsep}{1.2pt} 
    
    \definecolor{SoftBlue}{RGB}{235, 242, 250}

    \resizebox{\columnwidth}{!}{%
    \begin{tabular}{cl|ccc|ccc|ccc|c}
        \toprule
        \multirow{2}{*}{\textbf{Model}} & \multirow{2}{*}{\textbf{Method}} & \multicolumn{3}{c|}{\textbf{BadNet}} & \multicolumn{3}{c|}{\textbf{Blend}} & \multicolumn{3}{c|}{\textbf{WaNet}} & \textbf{Avg} \\
        \cmidrule(lr){3-5} \cmidrule(lr){6-8} \cmidrule(lr){9-11}
         & & 
         $\Delta_{BSR}$ & $\Delta_{ACC}$ & \textbf{Score} & 
         $\Delta_{BSR}$ & $\Delta_{ACC}$ & \textbf{Score} & 
         $\Delta_{BSR}$ & $\Delta_{ACC}$ & \textbf{Score} & 
         \textbf{Score} \\
        \midrule

         & CARE      & -97.90 & -13.42 & 84.48 & -98.18 & -5.43 & 92.75 & -97.75 & 1.56  & \textbf{99.31} & 92.18 \\
         & AI-Lancet & -100.00 & -71.64 & 28.36 & -99.80 & -83.30 & 16.50 & -92.38 & 1.12  & 93.50 & 46.12 \\
         & IR        & -97.49 & -17.96 & 79.53 & -96.06 & -27.70 & 68.36 & -100.00 & -4.02 & 95.98 & 81.29 \\
         & APRNN     & -98.20 & -10.29 & 87.91 & -99.90 & -1.66 & \textbf{98.24} & -92.63 & 0.67  & 93.30 & 93.15 \\
         & INNER     & -92.79 & -9.28  & 83.51 & -96.26 & -14.50 & 81.76 & -82.13 & -0.22 & 81.91 & 82.39 \\
         & SAU       & -68.35 & -15.94 & 52.41 & -52.74 & -19.80 & 32.94 & -89.58 & -10.30 & 79.28 & 54.88 \\
        \rowcolor{SoftBlue} 
        \cellcolor{white}\multirow{-7}{*}{\rotatebox{90}{\textbf{VGG13}}} 
         & \textbf{\tool} & -99.12 & -0.60 & \textbf{98.52} & -98.08 & -0.78 & 97.30 & -98.97 & -1.48 & 97.49 & \textbf{97.77} \\
        \midrule

         & CARE      & -100.00 & -5.60 & 94.40 & -83.18 & -5.88 & 77.30 & -100.00 & -1.43 & 98.57 & 90.09 \\
         & AI-Lancet & -87.50 & -9.93 & 77.57 & -30.84 & -5.34 & 25.50 & 16.31  & 0.20  & -16.10 & 28.99 \\
         & IR        & -99.39 & -24.00 & 75.39 & -96.64 & -5.45 & 91.19 & -100.00 & 0.51  & 100.50 & 89.03 \\
         & APRNN     & -98.63 & -5.17 & 93.46 & -99.81 & -31.60 & 68.21 & -100.00 & -28.20 & 71.80 & 77.82 \\
         & INNER     & -98.63 & 19.86 & \textbf{118.50} & -85.42 & -4.91 & 80.51 & -83.93 & -0.82 & 83.11 & 94.04 \\
         & SAU       & -91.29 & -5.24 & 86.05 & -99.77 & -31.10 & 68.67 & -99.86 & -60.90 & 39.04 & 64.59 \\
        \rowcolor{SoftBlue} 
        \cellcolor{white}\multirow{-7}{*}{\rotatebox{90}{\textbf{ResNet18}}}
         & \textbf{\tool} & -99.94 & 7.02 & 106.90 & -99.99 & 0.48 & \textbf{100.40} & -99.98 & 6.51 & \textbf{106.50} & \textbf{104.60} \\
        \midrule

         & CARE      & 0.00   & -2.74 & -2.74 & -1.75  & -1.69 & 0.06  & -0.75  & -2.14 & -1.39 & -1.36 \\
         & AI-Lancet & -99.80 & -84.00 & 15.80 & -97.94 & -52.80 & 45.14 & -96.50 & -80.50 & 16.00 & 25.65 \\
         & IR        & -88.83 & -38.30 & 50.53 & -96.40 & -52.40 & 44.00 & -87.86 & -40.40 & 47.46 & 47.33 \\
         & APRNN     & -88.23 & -0.12 & 88.11 & -80.88 & 1.33  & 82.21 & -81.66 & -2.58 & 79.08 & 83.13 \\
         & INNER     & -85.41 & -4.05 & 81.36 & -75.33 & -5.66 & 69.67 & -80.65 & -5.48 & 75.17 & 75.40 \\
         & SAU       & -86.06 & 5.54  & \textbf{91.60} & -34.24 & 0.94  & 35.18 & -96.29 & 6.62  & \textbf{102.90} & 76.56 \\
        \rowcolor{SoftBlue} 
        \cellcolor{white}\multirow{-7}{*}{\rotatebox{90}{\textbf{VGG16}}}
         & \textbf{\tool} & -99.84 & -11.30 & 88.54 & -86.94 & -2.62 & \textbf{84.32} & -88.94 & -1.98 & 86.96 & \textbf{86.61} \\
        \bottomrule
    \end{tabular}%
    }
    \vspace{-10pt}
\end{table}

\begin{table}[t]
    \centering
    \caption{Time cost comparison of different backdoor removal methods (measured in seconds).}
    \label{tab:time}
    \vspace{-10pt}  
    
    \scriptsize
    \renewcommand{\arraystretch}{1.1}
    \setlength{\tabcolsep}{3.5pt} 
    \definecolor{SoftBlue}{RGB}{235, 242, 250}

    \resizebox{\columnwidth}{!}{%
    \begin{tabular}{c | cc | cc | cc | c}
        \toprule
        \multirow{2}{*}{\textbf{Method}} & \multicolumn{2}{c|}{\textbf{VGG13-CIFAR10}} & \multicolumn{2}{c|}{\textbf{ResNet18-GTSRB}} & \multicolumn{2}{c|}{\textbf{VGG16-ImageNet10}} & \textbf{Avg} \\
        \cmidrule(lr){2-3} \cmidrule(lr){4-5} \cmidrule(lr){6-7} \cmidrule(lr){8-8}
         & Locating(s) & Repair(s) & Locating(s) & Repair(s) & Locating(s) & Repair(s) & \textbf{Total(s)} \\
        \midrule
        \care   & 43   & 466  & 46   & 543  & 34324 & 5543 & 13655 \\
        \ai     & 3228 & 0    & 3567 & 0    & 811   & 0    & 2535.3 \\
        \ir     & 4    & 8    & 20   & 19   & 1196  & 305  & 517.3 \\
        \aprnn  & 0    & 315  & 0    & 189  & 0     & 1447 & 650.3 \\
        \inner  & 226  & 146  & 315  & 154  & 389   & 229  & 486.3 \\
        \rowcolor{SoftBlue} 
        \textbf{\tool} & 7.55 & 23.95 & 4.27 & 23.5 & 12.32 & 168.8 & \textbf{80.1} \\
        \bottomrule
    \end{tabular}%
    }
    \vspace{-10pt} 
\end{table}

In Table \ref{tab:BSR}, \tool shows strong effectiveness in neutralizing diverse attack patterns (BadNet, Blend, WaNet), achieving the best average Score for each model family. Specifically, on VGG13, \tool reduces the Backdoor Success Rate (BSR) by an average of 98.73\% while incurring an average accuracy drop of 0.95\%. Performance on ResNet18 is notable: \tool achieves a very large average BSR reduction of 99.97\% and simultaneously \textit{improves} clean model accuracy by an average of 4.67\%. This trade-off is better than that of baselines such as AI-Lancet and IR, which often incur substantial accuracy loss (e.g., AI-Lancet degrades accuracy by 71.64\% on VGG13 BadNet) to achieve comparable repair.

Furthermore, \tool exhibits high computational efficiency. We report the locating time, repair time, and the average total time cost in Table \ref{tab:time}. By constraining the search space via Shapley-guided localization, \tool requires the lowest average total time (80.1 seconds) among all baseline methods. For instance, on the high-dimensional VGG16 model, \tool completes the entire repair process in approximately 181 seconds, whereas naïve search-based heuristics like CARE require over 39,000 seconds for the same task.

Similarly, in the task of adversarial mitigation (Table \ref{tab:ASR}), \tool achieves the highest average Score between robustness and accuracy. The framework reduces the Adversarial Success Rate (ASR) by 95.00\% on the VGG13 model and by 99.80\% on the ResNet18 model, with negligible accuracy costs of 0.66\% and 0.37\%, respectively. In contrast to methods such as \ir and \aprnn, which suffer from large accuracy losses (dropping by as much as 14.38\% and 33.98\% respectively), \tool successfully maintains model performance.

\begin{table}[t]
    \centering
    \caption{Effectiveness of \tool compared to baselines on adversarial repair.}
    \label{tab:ASR}
    \vspace{-10pt}
    
    \small 
    \renewcommand{\arraystretch}{1.05}
    \setlength{\tabcolsep}{3pt} 
    
    \definecolor{SoftBlue}{RGB}{235, 242, 250}

    \resizebox{\columnwidth}{!}{%
    \begin{tabular}{l | ccc | ccc | c}
        \toprule
        \multirow{2}{*}{\textbf{Method}} & \multicolumn{3}{c|}{\textbf{VGG13-CIFAR10}} & \multicolumn{3}{c|}{\textbf{ResNet18-GTSRB}} & \textbf{Avg} \\
        \cmidrule(lr){2-4} \cmidrule(lr){5-7} \cmidrule(l){8-8} 
         & $\Delta_{ASR}\%$ & $\Delta_{ACC}\%$ & \textbf{Score} & $\Delta_{ASR}\%$ & $\Delta_{ACC}\%$ & \textbf{Score} & \textbf{Score} \\
        \midrule
        
        \care   & -90.10 & -2.05 & 88.05 & -96.20 & -2.04  & 94.16 & 91.11 \\
        \ai     & -59.50 & -1.26 & 58.24 & -27.90 & -1.92  & 25.98 & 42.11 \\
        \ir     & -99.90 & -14.38& 85.52 & -86.20 & -7.93  & 78.27 & 81.90 \\
        \aprnn  & -68.60 & -0.57 & 68.03 & -96.70 & -33.98 & 62.72 & 65.38 \\
        \inner  & -72.20 & -0.11 & 72.09 & -99.50 & -10.19 & 89.31 & 80.70 \\
        \piat   & -82.37 & 0.00  & 82.37 & -73.45 & -3.16  & 70.29 & 76.33 \\
        
        \rowcolor{SoftBlue} 
        \textbf{\tool} 
        & -95.00 & -0.66 & \textbf{94.34} 
        & -99.80  & -0.37 & \textbf{99.43} 
        & \textbf{96.89} \\
        \bottomrule
    \end{tabular}%
    }
    \vspace{-10pt}
\end{table}

\begin{table*}[t]
    \centering
    \caption{Effectiveness of \tool compared to baselines on fairness repair.}
    \label{tab:UF}
    \vspace{-5pt}

    \small 
    \renewcommand{\arraystretch}{1.05}
    \setlength{\tabcolsep}{4pt} 

    \definecolor{SoftBlue}{RGB}{235, 242, 250}

    \resizebox{\textwidth}{!}{%
    \begin{tabular}{ll | ccc | ccc | ccc | ccc | ccc}
    \toprule

    \multirow{2}{*}{\textbf{Dataset}} & \multirow{2}{*}{\textbf{Feat.}} & 
    \multicolumn{3}{c|}{\textbf{CARE}} & \multicolumn{3}{c|}{\textbf{PRDNN}} & 
    \multicolumn{3}{c|}{\textbf{ADF}} & \multicolumn{3}{c|}{\textbf{IDNN}} & \multicolumn{3}{c}{\textbf{\tool}} \\
    \cmidrule(lr){3-5} \cmidrule(lr){6-8} \cmidrule(lr){9-11} \cmidrule(lr){12-14} \cmidrule(l){15-17}
    & & $\Delta_{UF}\%$ & $\Delta_{ACC}\%$ & \textbf{Score} & 
    $\Delta_{UF}\%$ & $\Delta_{ACC}\%$ & \textbf{Score} & 
    $\Delta_{UF}\%$ & $\Delta_{ACC}\%$ & \textbf{Score} & 
    $\Delta_{UF}\%$ & $\Delta_{ACC}\%$ & \textbf{Score} & 
    $\Delta_{UF}\%$ & $\Delta_{ACC}\%$ & \textbf{Score} \\
    \midrule

    Census & Race   & -70.67 & -3.49 & 67.18 & -23.39 & -1.71 & 21.68 & -47.52 & -1.84 & 45.68 & -79.39 & -4.37 & 75.02 & -90.80 & -2.37 & \textbf{88.43} \\
    Census & Gender & -35.94 & -3.13 & 32.81 & -12.67 & -1.43 & 11.24 & -43.23 & -2.03 & 41.20 & -72.22 & -4.53 & \textbf{67.69} & -57.14 & -0.28 & 56.86 \\
    Census & Age    & -47.24 & -4.05 & 43.19 & -32.90 & -3.47 & 29.43 & -39.11 & -2.50 & 36.61 & -73.16 & -4.91 & 68.25 & -79.09 & -1.45 & \textbf{77.64} \\
    Credit & Gender & -21.74 & -0.84 & 20.90 & -11.10 & -0.55 & 10.55 & -31.53 & -1.87 & 29.66 & -54.76 & -2.68 & 52.08 & -73.83 & -3.06 & \textbf{70.77} \\
    Credit & Age    & -45.17 & -0.57 & 44.60 & 4.05   & -1.97 & -6.02 & -36.04 & -2.76 & 33.28 & -62.25 & -3.36 & 58.89 & -89.40 & -2.04 & \textbf{87.36} \\
    \midrule

    \rowcolor{SoftBlue}
    \multicolumn{2}{c|}{\textbf{Average}} & 
    -44.15 & -2.42 & 41.73 & 
    -15.20 & -1.83 & 13.37 & 
    -39.49 & -2.20 & 37.29 & 
    -68.36 & -3.97 & 64.39 & 
    -78.05 & -1.84 & \textbf{76.21} \\

    \bottomrule
    \end{tabular}%
    }
    \vspace{-10pt}
\end{table*}

Finally, the framework's effectiveness extends to fairness repair (Table \ref{tab:UF}). Following the unified trade-off evaluation principle, \tool achieves the best average Score (76.21\%) among all baselines. It decreases unfairness by an average of 78.05\%, with a correspondingly minimal average accuracy decrease of only 1.84\%. While \idnn performs marginally better on the specific Census (Gender) attribute, \tool consistently dominates the remaining four tasks and the overall average. For instance, compared to the next-best method (\idnn), \tool achieves a 9.69\% greater average unfairness reduction while incurring less than half the accuracy loss (1.84\% vs. 3.97\%). This result underscores \tool's ability to correct defects while preserving utility.

In general, \tool surpasses the best baselines by average margins of up to 10.56\% on backdoor removal (compared to INNER), 5.78\% on adversarial mitigation (compared to CARE), and 11.82\% on unfairness repair (compared to \idnn). Notably, compared to specialized methods such as SAU \cite{DBLP:conf/nips/WeiZZW23} and PIAT \cite{DBLP:journals/corr/abs-2303-13955}, \tool often keeps the accuracy drop small, though some settings still exceed 2\%, whereas baselines frequently suffer severe utility loss.

\subsubsection{RQ3: What factors influence \tool's performance?}\hspace*{\fill}
\label{rq3}

\begin{figure}[t]
    \centering
    \includegraphics[width=0.75\linewidth]{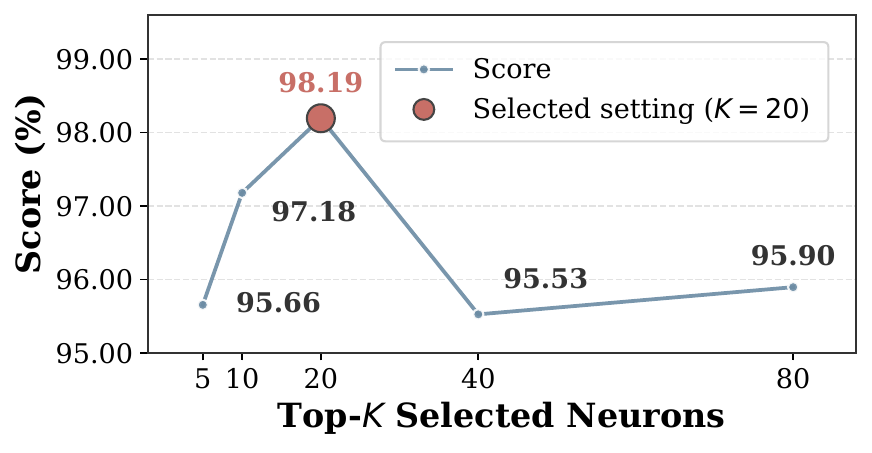}
    \vspace{-8pt}
    \caption{Effect of the top-$K$ neurons on repair performance for \textit{VGG13-CIFAR10-BadNets}.}
    \label{fig:rq3_topk}
    \vspace{-12pt}
\end{figure}

\begin{figure}[t]
    \centering
    \includegraphics[width=\linewidth]{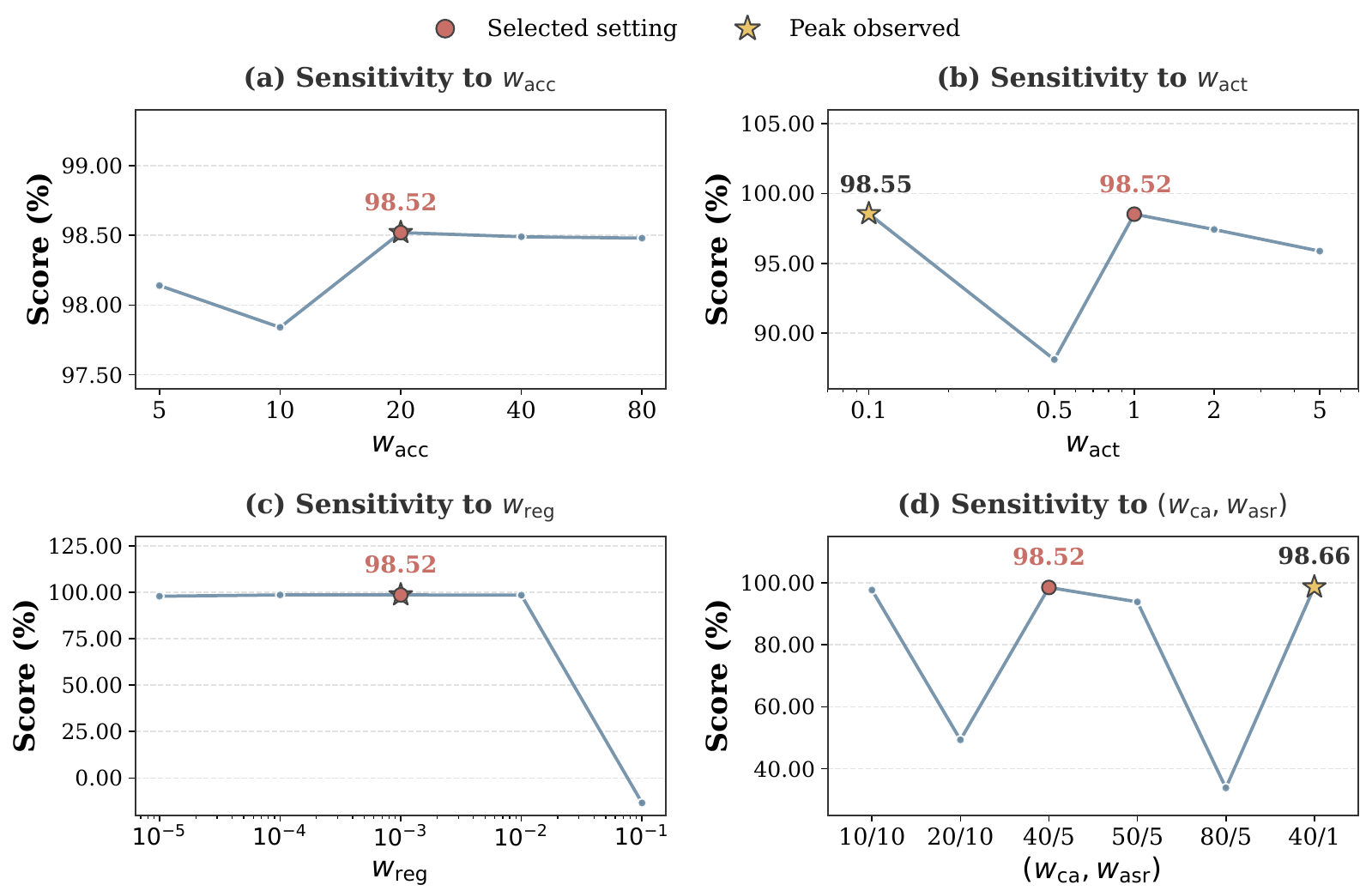}
    \vspace{-15pt}
    \caption{Sensitivity analysis of the repair objective weights on \textit{VGG13-CIFAR10-BadNets}.}
    \label{fig:rq3_weight_sensitivity}
    \vspace{-16pt}
\end{figure}

The performance of \tool is inherently governed by key hyper-parameters during its two primary stages: fault localization and neuron repair. To evaluate the impact of hyper-parameters, we conduct sensitivity analysis using the \textit{VGG13-CIFAR10-BadNets} configuration as a representative case.

In the localization stage, the number of selected neurons ($K$) controls the essential trade-off between repair capacity and search space complexity. As illustrated in Figure~\ref{fig:rq3_topk}, the overall score $S_c$ peaks at $K=20$ and subsequently declines. This trend indicates that an overly small $K$ fails to encompass all repair-critical neurons, whereas an excessively large $K$ introduces unnecessary parameter perturbations that complicate the black-box optimization. Notably, empirical evaluation confirms that the choice of $K$ does not significantly impact the overall computational overhead. This efficiency stems from the fact that the time complexity of the derivative-free optimization is dominated by the forward passes required for fitness evaluation, which remains relatively constant regardless of the localized weight space's dimensionality.

During the neuron repair stage, the optimization trajectory is directed by the weights assigned to the objective function, which precisely balances benign accuracy preservation ($w_{\text{acc}}$, $w_{\text{ca}}$), property violation mitigation ($w_{\text{asr}}$), activation deviation ($w_{\text{act}}$), and weight modification constraints ($w_{\text{reg}}$). Figure~\ref{fig:rq3_weight_sensitivity} presents a one-factor-at-a-time (OFAT) sensitivity analysis, where each parameter is individually varied while maintaining the default configuration $(w_{\text{acc}}, w_{\text{act}}, w_{\text{reg}}, w_{\text{ca}}, w_{\text{asr}}) = (20, 1, 0.001, 40, 5)$. The analysis reveals that the selected default setting consistently achieves best-or-near-best performance across all parameter sweeps. This robust performance aligns perfectly with our design intuition: heavily weighting clean accuracy ($w_{\text{acc}}$) and clean activation ($w_{\text{ca}}$) ensures the repaired model strictly adheres to the original pre-trained manifold, effectively preventing catastrophic forgetting. Conversely, a minimal regularization weight ($w_{\text{reg}}$) provides just enough constraint to avoid unbounded parameter drift without stifling the optimizer's ability to erase the defect. Furthermore, execution logs demonstrate that tuning these objective weights does not introduce computational bottlenecks. The total repair time exhibits minimal fluctuation, highlighting the stability of the underlying optimization process.

\subsubsection{RQ4: How useful is Shapley-guided fault localization?}\hspace*{\fill}
\label{rq4}

To evaluate Shapley-guided fault localization, we investigate its localization accuracy and generality across optimizers.

\begin{table}[t]
    \centering
    \caption{Ablation on localization strategies.}
    \label{tab:loc_abl}
    \vspace{-5pt}
    
    \scriptsize 
    \renewcommand{\arraystretch}{1.1} 
    \setlength{\tabcolsep}{4pt} 

    \resizebox{\columnwidth}{!}{%
    \begin{tabular}{@{} l ccc ccc ccc @{}}
        \toprule
        \multirow{2}{*}{\textbf{Task}} & \multicolumn{3}{c}{\textbf{Random}} & \multicolumn{3}{c}{\textbf{SHARPEN (Bottom-$k$)}} & \multicolumn{3}{c}{\textbf{SHARPEN (Top-$k$)}} \\
        \cmidrule(lr){2-4} \cmidrule(lr){5-7} \cmidrule(l){8-10}
        & $\Delta_{Prop}$ & $\Delta_{ACC}$ & $S_c$ & $\Delta_{Prop}$ & $\Delta_{ACC}$ & $S_c$ & $\Delta_{Prop}$ & $\Delta_{ACC}$ & $S_c$ \\
        \midrule
        BadNets-VGG13 & -0.51  & -1.53 & -1.02 & -0.05  & -0.14 & -0.09 & -99.12 & -0.60 & \textbf{98.52} \\
        FGSM-VGG13    & -0.48  & -0.66 & -0.18 & -0.66  & -0.06 & 0.60  & -95.00 & -0.66 & \textbf{94.34} \\
        Race-Census   & -36.36 & -1.54 & 34.82 & -16.67 & -3.05 & 13.62 & -90.80 & -2.37 & \textbf{88.43} \\
        \bottomrule
    \end{tabular}%
    }
    \vspace{-8pt}
\end{table}
\textit{Localization Effectiveness.} We compare \tool (\textit{Top-$k$} suspicious neurons) with two baselines: targeting $k$ random neurons (\textit{Random}) and targeting the least suspicious neurons (\textit{Bottom-$k$}). As shown in Table \ref{tab:loc_abl}, optimizing random or bottom-ranked neurons does not consistently mitigate property violations and yields substantially lower $S_c$ scores than the Top-$k$ strategy. These results suggest that randomly selected or low-ranked neurons are less relevant to the faulty behavior. Moreover, perturbing the \textit{Bottom-$k$} neurons can reduce accuracy without yielding meaningful property repair, indicating that these neurons are more important for normal task performance than for the defect itself. In contrast, \tool identifies neurons that are more closely related to the defect, leading to substantial property reductions and the highest $S_c$ scores across all three tasks ($>88\%$). This result suggests that effective repair strongly depends on accurate fault localization.
\begin{table}[t]
    \centering
    \caption{Comparison of fairness and accuracy percentage changes using different derivative-free optimizers.}
    \label{tab:UF_abl}
    \vspace{-5pt} 
    
    \small 
    \renewcommand{\arraystretch}{1.05}
    \setlength{\tabcolsep}{6pt} 
    \definecolor{SoftBlue}{RGB}{235, 242, 250}

    \begin{tabular}{ll | cc | cc}
        \toprule
        \multirow{2}{*}{\textbf{Dataset}} & \multirow{2}{*}{\textbf{Feat.}} & \multicolumn{2}{c|}{\textbf{\tool-CMA}} & \multicolumn{2}{c}{\textbf{\tool-RACOS}} \\
        \cmidrule(lr){3-4} \cmidrule(l){5-6}
        & & $\Delta_{UF}\%$ & $\Delta_{ACC}\%$ & $\Delta_{UF}\%$ & $\Delta_{ACC}\%$ \\
        \midrule
        Census & Race   & -90.80 & -2.37 & -93.51 & -4.80 \\
        Census & Gender & -57.14 & -0.28 & -97.22 & -3.33 \\
        Census & Age    & -79.09 & -1.45 & -68.42 & -0.85 \\
        Credit & Gender & -73.83 & -3.06 & -80.33 & -2.04 \\
        Credit & Age    & -89.40 & -2.04 & -97.51 & -4.08 \\
        \midrule
        \rowcolor{SoftBlue}
        \multicolumn{2}{c|}{\textbf{Average}} & \textbf{-78.05} & \textbf{-1.84} & \textbf{-87.40} & \textbf{-3.02} \\
        \bottomrule
    \end{tabular}
    \vspace{-12pt}
\end{table}

\textit{Generality Across Optimizers.} A robust localization method should identify critical neurons that can be effectively used with different repair algorithms. To examine this, we replaced CMA-ES with RACOS while repairing the same neurons localized by \tool, using fairness repair as a representative case (Table \ref{tab:UF_abl}). Both optimizers achieve substantial unfairness reductions (87.40\% for RACOS vs. 78.05\% for CMA-ES), suggesting that repair effectiveness is not tied to a specific optimizer and is instead largely enabled by accurate neuron identification. Although RACOS reduces unfairness slightly more, it incurs greater accuracy degradation (-3.02\% vs. -1.84\%), making CMA-ES a better default choice in terms of the property--accuracy trade-off. This difference likely arises from their distinct search mechanisms. RACOS, a randomized coordinate shrinking algorithm, searches within shrinking bounding boxes to satisfy the target constraint, which may disrupt the original decision boundaries more strongly. In contrast, CMA-ES adapts its search distribution through a covariance matrix, resulting in more conservative parameter updates that better preserve the utility of the pre-trained model.

\section{Related Work}
\label{sec:related}

The methods for neural network repair are broadly divided into two categories: \textit{Data-Driven Repair} and \textit{Model-Internal Repair}. Data-driven approaches repair model behavior by retraining on augmented datasets. For instance, \adf \cite{DBLP:conf/icse/ZhangW0D0WDD20} generates discriminatory samples to improve fairness, \deepr \cite{DBLP:journals/tr/YuQGJXMZ22} uses style-guided data augmentation to incorporate unknown failure patterns, and \fsgm \cite{DBLP:conf/icsm/RenYQJLXMZ20} generates new data based on the distribution of known error samples. However, these methods are often computationally intensive and do not guarantee repair of specific defects. Consequently, research has shifted towards model-internal repair, which modifies a subset of a pre-trained model's parameters, typically via fault localization followed by repair execution. Fault localization techniques have evolved from statistical methods to more principled approaches. Early methods relied on gradients or influence functions, as seen in \ir \cite{DBLP:conf/sac/HenriksenLL22} and the model-based analysis in \rnnr \cite{DBLP:conf/icml/Xie0MLWZLX21} for recurrent networks, whereas later methods such as \ai \cite{DBLP:conf/ccs/0018Z0Z21} use differential feature analysis. Other techniques include causality analysis in \care \cite{DBLP:conf/icse/Sun0PS22}, interpretability probes in \inner \cite{DBLP:conf/issta/ChenZSWXY24}, behavior imitation in \birdnn \cite{DBLP:journals/corr/abs-2305-03365}, and formal verification in \vere \cite{DBLP:conf/icse/MaYWSHW24}. Specialized methods for fairness, such as \fairn \cite{DBLP:conf/icse/GaoZMSCW22} and \neuronf \cite{DBLP:conf/icse/ZhengCD0CJW0C22}, identify and retrain neurons responsible for biased behavior. The repair execution phase also employs diverse strategies. Search-based methods use heuristic algorithms to modify identified faulty neurons. For instance, \arachne \cite{DBLP:journals/tosem/SohnKY23} uses differential evolution, while \care and \birdnn utilize particle swarm optimization. A distinct category of repair relies on formal methods and constraint solving for stronger guarantees. Works like \prdnn \cite{DBLP:conf/pldi/SotoudehT21}, \aprnn \cite{DBLP:journals/pacmpl/TaoNMT23}, \nnr \cite{DBLP:conf/cav/UsmanGSNP21}, and \reassure \cite{DBLP:conf/iclr/FuL22} formulate the repair task as a Linear Programming (LP) or constraint satisfaction problem. This approach is further advanced by \autoric \cite{DBLP:journals/tosem/SunLWCTM25}, which formulates the objective as a Quadratic Programming (QP) problem solved via a constrained optimizer. Finally, another strategy is direct structural modification, where frameworks like \idnn \cite{DBLP:conf/issta/ChenWS0024} ``isolate'' faulty neurons by fixing their outputs rather than altering weights. To evaluate these diverse techniques, benchmarking platforms like \airepair \cite{DBLP:conf/icse/SongSMC23} provide modular support for evaluating repair methods.

\vspace{-0.5mm}
\section{Conclusion and Future Work}
\label{sec:conclusion}

In this paper, we present \tool, a derivative-free repair framework that integrates SHAP-based interpretable fault localization with CMA-ES to mitigate backdoors, adversarial vulnerabilities, and unfairness in DNNs. Experimental results across multiple tasks show that \tool provides a favorable balance between repair effectiveness, utility preservation, and computational efficiency. These findings suggest that combining interpretable localization with derivative-free optimization is a promising direction for trustworthy model repair.

A current limitation of \tool lies in the computational cost of Shapley estimation, especially for large-scale foundation models and complex generative settings. Future work will focus on improving the scalability of fault localization through sparsity-aware sampling and extending the framework to broader model families, including Large Language Models (LLMs).

\clearpage

\bibliographystyle{ACM-Reference-Format}
\bibliography{ref.bib}

\end{document}